\documentclass[12pt]{article}
\input{psfig.sty}
\usepackage{color}

\def\mydate{26th August 2000}
\def\ignore#1{{}}

\newcommand{\athird}{\frac{1}{3}}
\newcommand{\twothird}{\frac{2}{3}}

\newcommand{\beeq}{\begin{equation}}
\newcommand{\eneq}{\end{equation}}
\newcommand{\be}{\begin{eqnarray}}
\newcommand{\ee}{\end{eqnarray}}

\def\ppr{\vec{p^{\prime}}}
\def\pp{\vec{p}}

\def\dd{\partial}
\def\ppr{\vec{p^{\prime}}}
\def\pp{\vec{p}}

\def\la{\raise.16ex\hbox{$\langle$}\lower.16ex\hbox{}  }
\def\ra{\, \raise.16ex\hbox{$\rangle$}\lower.16ex\hbox{} }

\def\psibar{ \psi \kern-.65em\raise.6em\hbox{$-$} \lower.6em\hbox{} }
\def\psibaralpha{ \psi^{(\alpha)} \kern-1.9em\raise.6em\hbox{$-$}
\kern+1.2em\hbox{}}
\def\psibara{ \psi^{(a)} \kern-1.9em\raise.6em\hbox{$-$}\kern+1.2em\hbox{}}

\begin{document}

\baselineskip=12pt

{\small \noindent \mydate \hfill NUC-MINN-00/15--T}

{\small  \hfill BNL-NT-00/21}

\baselineskip=40pt plus 1pt minus 1pt

\vskip 3cm

\begin{center}

{\Large\bf From colored glass condensate to gluon plasma: 
equilibration in high energy heavy ion collisions}\\

\vspace{1.0cm}
\baselineskip=20pt plus 1pt minus 1pt

{\large  Jefferson Bjoraker$^1$ and  Raju Venugopalan$^{1,2}$}\\
\vspace{.1cm}
{\it \small $^1$ Physics Department, Brookhaven National Laboratory,}
{\it  Upton, NY. 11973, U.S.A.}\\
{\it \small $^2$ RIKEN-BNL Research Center, Brookhaven National Laboratory,}
{\it  Upton, NY. 11973, U.S.A.}
\\ 
\end{center}

\vskip 2.cm
\baselineskip=20pt plus 1pt minus 1pt

\begin{abstract}

The initial distribution of gluons at the very early times after a high energy
heavy ion collision is described by the bulk scale $Q_s$ of gluon
saturation in the nuclear wavefunction. The subsequent
evolution of the system towards kinetic equilibrium is described by 
a non--linear Landau equation for the single
particle distributions~\cite{Mueller1,Mueller2}. In this paper, 
we solve this equation numerically for the 
idealized initial conditions proposed by
Mueller, and study the evolution of the system
to equilibrium.  We discuss the sensitivity of our results on the 
dynamical screening of collinear divergences. In a particular
model of dynamical screening, the convergence to the hydrodynamic limit is 
seen to be rapid relative to hydrodynamic time scales. The
equilibration time, the initial temperature, and the chemical
potential are shown to have a strong functional dependence on the initial 
gluon saturation scale $Q_s$.
 
\end{abstract}

\section{Introduction}
\vskip 0.1in

An outstanding problem in high energy nuclear scattering is whether the 
hot and dense matter formed equilibrates to briefly form a plasma of 
deconfined quarks and gluons--the quark gluon plasma. This problem is of 
great topical interest, with collisions already taking place at the 
Relativistic Heavy Ion Collider (RHIC) and scheduled to take place several 
years hence at the Large Hadron Collider (LHC).

Whether or not a quark gluon plasma is formed depends strongly on 
the highly non--equilibrium initial distributions of partons formed 
immediately after the collision. Clearly, these distributions must 
influence the subsequent evolution of the system towards equilibrium. 
Furthermore, the problem is complicated by the rapid expansion of the 
system as a whole. The 
magnitude of the collision induced relaxation time 
relative to the expansion time 
is what determines whether equilibrium is indeed reached.

The study of equilibration in relativistic heavy ion collisions is as
old as the subject itself. However, very few {\it ab
initio} studies exist that attempt to follow the evolution of the
system, all the way, from the first instants of the collision to
equilibrium~\cite{Geiger,Karietal, Wang}. This was so because it was not
known how to treat, in a self--consistent manner, the small Bjorken $x$
``wee'' parton modes that are responsible for particle production at
central rapidities. These small $x$ modes provide the initial conditions 
for the space--time evolution of the partonic matter formed in heavy 
ion collisions.

Because the occupation number of small $x$ modes in the nuclear 
wavefunctions is large, it was shown that classical methods could be 
used to compute their distributions~\cite{MV}. This classical effective 
theory is, in exact analogy to a spin glass, a color glass 
condensate~\cite{Larry,RajGavai}, and is characterized by a bulk scale 
$Q_s$ -- the momentum scale at which gluon distributions 
saturate. Analytical expressions for the parton distributions were 
obtained in Refs.~\cite{JKMW} and \cite{Kovchegov}.
Subsequently, quantum corrections 
to the classical ``non-Abelian Weizs\"acker--Williams'' fields 
were computed~\cite{AJMV}, and renormalization group methods~\cite{JKMW,JKLW} 
devised to 
study how the classical parton distributions in the nuclei 
changed with energy (or equivalently, with $x$). The scale $Q_s$ now 
depends on $x$. This dependence is represented by 
a line in the $x$--$Q^2$ plane -- it separates the 
saturated, non--linear regime 
of QCD at high parton densities from that of linear QCD evolution. That 
$Q_s$ is a function of $x$, and grows as one goes to smaller $x$, or 
equivalently, to higher energies, will only be implicit in this work.

In the classical effective field theory approach, the problem of
initial conditions can be formulated as the problem of finding
solutions of the Yang--Mills equation with initial conditions given by
the classical fields of each of the nuclei {\it before} the
collision~\cite{KMW}. Since analytical expressions exist for the classical 
fields of the nuclei before the collision, the initial conditions are 
fully determined.

Perturbative solutions of the Yang--Mills equations, which describe
classical gluon production to lowest order, have been discussed by
several authors~\cite{KMW,PYM}. These were found to be infra--red
divergent. Within the Yang--Mills approach, a fully non--perturbative
treatment is therefore necessary. Non--perturbative, numerical
solutions of the Yang--Mills equations have been found
recently~\cite{AlexRaj1,AlexRaj2}.  In particular, one now
knows the initial number and energy distribution of gluons after a
collision~\cite{AlexRaj3}. An idealized form of the initial gluon
distribution, in terms of the saturation scale of the color glass
condensate $Q_s$, was given recently by Mueller~\cite{Mueller1}.
These idealized distributions are sufficient for the purposes of this
paper. We will reserve a more quantitative analysis using the initial
distributions of Ref.~\cite{AlexRaj3} for a future work~\cite{JeffRaj}.

The initial partonic system is completely out of
equilibrium. The subsequent scattering and evolution of the system
towards equilibrium was studied by
Mueller~\cite{Mueller1,Mueller2}. In Ref.~\cite{Mueller2}, he showed
that, under the assumption that the system is undergoing boost invariant 
expansion,  
the evolution of single particle distributions could be described
by a non--linear Landau equation. At very early times, this equation
can be linearized~\footnote{to a Fokker-Planck
equation~\cite{LifPit}.}, and studied analytically. However, the
analytical approximations soon break down, and the evolution of the
system cannot be followed analytically all the way to equilibrium.

In this paper, we numerically solve the Landau equation proposed by
Mueller.  (Along the way, we compare our results to Mueller's analytical 
results and show quantitatively where the analytical
approach breaks down.)  We are thus able to follow the evolution of
the system all the way to equilibrium. We study the dependence of the
equilibration time, the initial temperature, and the chemical
potential on the saturation scale $Q_s$ of the nuclear wavefunction,
and on $\alpha_S$. (Note: one may estimate $Q_s\sim 1$ GeV at RHIC and
$Q_s\sim 2-3$ GeV at LHC~\cite{Mueller1,Raj99}.) We discuss 
the dynamical screening of the collinear divergence arising from small angle
scattering.  Our results for equilibration are obtained, primarily, in
a particular model of dynamical screening~\cite{BiroMuller}.

To conclude whether local kinetic equilibrium is attained in a system
undergoing boost invariant 1+1--dimensional expansion, one should
compare the equilibration time $t_{equil}$ to the hydrodynamic
expansion time $t_{hydro}$. Typically, $t_{hydro}\approx R/c_0$, where
$R$ is the radius of the nucleus, and $c_0$ is the speed of sound in
the system. For an ideal ultrarelativistic gas, $c_0= 1/\sqrt{3}$.
Our qualitative results suggest that for the energies of interest at
RHIC and LHC, $t_{equil}\ll t_{hydro}$, thereby indicating that
favorable conditions may exist for the formation of a hot gluon plasma 
at RHIC and at LHC.

In studying the likelihood of equilibration, we have only considered
number conserving $2\longrightarrow 2$ processes. Naively,
$2\longrightarrow 3$ processes will be suppressed by a power of
$\alpha_S$, and may be considered sub--leading. However,
$2\longrightarrow 3$ processes may, in principle, be much more efficient than
$2\longrightarrow 2$ processes in driving the system towards 
thermal equilibrium~\cite{AMDScomm}. Whether this is indeed the case is a
dynamical question we will not address in this work. Regardless,
$2\longrightarrow 2$ processes should set an upper bound on the
equilibration time. Furthermore, it is likely that the numerical techniques 
developed here can be adapted to quantitatively study 
the effects of number changing processes.

The paper is organized as follows. We begin in section 2 by discussing
the initial conditions for gluon production in nuclear
collisions. After the collision, particle production is described by
the space--time evolution of classical gauge fields. At late times,
when the system is dilute, one can define partons, and study their
interactions in a kinetic approach. The kinetic transport equation
derived by Mueller is described in section 3.  In the following
section, we discuss the problem of dynamical screening of collinear
divergences, and describe the particular model of screening that is
employed in this work. In section 5, we first briefly outline the
numerical method, and compare results for a linearized
(Fokker--Planck) version of the Landau equation with Mueller's
analytical results. We next discuss numerical results from the
solution of the full Landau equation.  Various measures of
equilibration are described and computed as a function of the
saturation scale $Q_s$ and the coupling $\alpha_S$. We summarize our
results in the final section and comment on their experimental
ramifications. We will also discuss the various uncertainties that
need to be further quantified, and their likely impact on
equilibration in high energy heavy ion collisions. An appendix
contains a detailed description of the numerical method employed to
solve the Landau equation.

\section{The initial conditions for parton evolution}
\vskip 0.1in

The problem of initial conditions in heavy ion collisions is as
follows.  Before the collision, the nucleus, as viewed in a particular
gauge, is a coherent superposition of various Fock modes containing
differing numbers of partons~\footnote{For an illuminating exposition
of this point of view, see Ref.~\cite{BjKoSo}.}. Partons in Fock modes
containing a large number of partons, $|qqq\cdots qgggg\cdots
q\bar{q}\rangle$, have small values of $x$--in the language of the
parton model, the momentum of the nucleus is shared among a large
number of constituents. Each parton has a $+$ momentum $k^+$, and
$x=k^+/P^+$, where $P^+$ is the momentum of the nucleus. The large
multiplicities at central rapidities--which correspond to small
$x$--are obtained when these small $x$ Fock modes ``go on-shell''. One
should note that even though the small $x$ modes carry small
longitudinal momenta, their transverse momenta $p_t$ can be large.

These small $x$ partons are highly de--localized in the longitudinal $x^-$ 
direction~\footnote{if $P^+$ is the momentum of the nucleus} -- 
they have large wavelengths relative to the Lorentz 
contracted width of the nucleus. In addition, these virtual fluctuations 
are very short--lived. Any attempt to treat them in a naively classical 
transport approach is therefore problematic. An approach based on pQCD 
collinear factorization is also inadequate unless the partons also have 
very large transverse momenta. This is because such a picture is predicated 
on convolving the probabilities of parton distributions in the 
two nuclei with the elementary parton--parton scattering cross-section. 
It cannot describe coherence effects which are important at small $x$. 
(For large $p_t$, these effects are suppressed.)

It was realized some time ago that the small $x$ Fock states in nuclei
responsible for multi--particle production at central rapidities are
states of high occupation number. As a consequence, their
distributions in the nuclei are described by a classical effective
field theory (EFT)~\cite{MV}. The classical distributions in a single nucleus 
can be solved and an analytical form for the distributions 
obtained~\cite{JKMW,Kovchegov}. The classical gluon distribution falls off 
as $1/k_t^2$ at large transverse momentum $k_t$ but saturates at smaller 
$k_t$--growing only slowly as $\ln(\alpha_S\mu/k_t)/\alpha_S$. The infrared 
structure of the EFT is analogous to that of a spin glass in condensed matter 
systems~\cite{RajGavai}--it is thus a colored glass condensate~\cite{Larry}.

The EFT approach was used in Ref.~\cite{KMW} to 
treat the problem of initial conditions in nuclear collisions. It consists 
of solving the Yang--Mills equations in the forward light cone, after 
the collision, with initial conditions given by the (known) gauge fields of 
the nuclei before the collision.

Gluons at central rapidities are produced classically. (Boost invariance is 
assumed.)  Perturbative
computations of the transverse momentum distributions of classically
produced gluons (lowest order in $\alpha_S$ and lowest non--trivial
order in $\alpha_S Q_s/p_t$) were first computed in Ref.~\cite{KMW},
and later by several authors~\cite{PYM}. In these computations, the 
number distribution is found to be infra-red divergent. They agree with 
the lowest order 
mini--jet predictions at large $p_t$~\footnote{For detailed comparisons, 
 see the papers by Gyulassy and McLerran, and by Guo, in Ref.~\cite{PYM}}.

When $p_t\leq Q_s$, all orders in $Q_s/p_t$ contribute
equally. The perturbative computations, to lowest order in $Q_s/p_t$ are 
therefore not sufficient. Thus, even at lowest order in
$\alpha_S$, one needs to resum all orders in $Q_s/p_t$. An
analytic solution to this non--perturbative problem has not been
found. However, the problem was recently formulated as a
2+1--dimensional classical effective field theory and solved
numerically (for an SU(2) gauge theory) on a 2--dimensional transverse
lattice~\cite{AlexRaj1}. The evolution of gauge fields is
 computed in real time, and
at late times, the energy~\cite{AlexRaj2} and the
number~\cite{AlexRaj3} of produced gluons at central rapidities can be
computed. 
The initial distribution of gluons, from our numerical 
simulations~\cite{AlexRaj3}, is as follows. At low values of $p_t$, 
$p_t\leq Q_s$ it is remarkably like the Bose--Einstein distribution for a 
gas of massive particles in two dimensions. The gluons acquire a screening 
mass due to strong non--linear interactions -- this also renders the 
distributions infrared finite. At large $p_t$, $p_t \gg Q_s$, the 
distribution is a power law, $Q_s^4/p_t^4$.
The Born--pQCD prediction for the 
gluon distribution is therefore obtained, as expected, in 
the large $p_t$ limit.

The gluons produced at central rapidities are completely out of equilibrium. 
The study of the evolution of this system towards equilibrium was initiated 
by Mueller~\cite{Mueller1,Mueller2}. For simplicity, Mueller chose 
idealized  initial conditions--a constant instead of 
a Bose--Einstein distribution for $p_t\leq Q_s$, 
and zero for $p_t\gg Q_s$--namely, the 
theta function $\theta(Q_s^2-p_t^2)$. In this work, we will consider only 
these idealized distributions of Mueller. In a later work~\cite{JeffRaj}, 
we will consider 
more realistic distributions computed very recently 
from the lattice simulations of 
Ref.~\cite{AlexRaj3}. In general, the initial spatial distribution 
per unit rapidity is
\be
{1\over L^2}\,{dN\over d\eta} = c\,{N_c^2-1\over {4\pi^2\alpha_S N_c}}\,
Q_s^2\,\,,
\ee
where $L^2$ is the transverse area, $\eta$ is the space--time rapidity, 
and $c$, determined non--perturbatively,  is a weak function of $Q_s R$. 
For an SU(2) gauge theory, it was 
computed recently in Ref.~\cite{AlexRaj3} to be $c= 1.29\pm 0.09$ in the 
regime of interest.

The initial distributions are highly anisotropic, with the produced partons 
having zero longitudinal momentum $p_z$. Assuming boost invariance, and 
using the relation
\be
f(x,p) = {(2\pi)^3\over {2(N_c^2-1)}}\,{dN\over d^3p\, d^3x} \,,
\ee
for the single particle distribution $f(x,p)$, one finds at late times 
that~\footnote{The $\delta$--function distribution is of order $1/\Delta p_z$.
We shall see later, from the early time analytical 
solution of the Fokker--Planck equation, that the 
longitudinal momentum changes little in the time in takes for 
the occupation number to decrease below unity.}
\be
f(x,p) = {c\over \alpha_S N_c}\, {1\over t}\,\delta(p_z)\,
\theta(Q_s^2-p_t^2)\, .
\label{idis}
\ee
At late times, the occupation number of partons in the transverse plane 
becomes small, and their evolution can no longer be followed 
on the lattice. However, when the occupation number is not too large, 
transport theory can be applied to study the further evolution of the system.
The initial condition for this evolution is given by the single particle 
distribution in Eq.~\ref{idis}.

Using the results of Ref.~\cite{AlexRaj2}, which determined the time
at which gluons come on-shell after a collision, we can estimate the
time at which small angle scattering between the gluons can be described by
the Boltzmann equation.  The formation
times, using $Q_s = 1$ GeV for RHIC and $Q_s = 2$--$3$ GeV for LHC, are
\begin{equation}
t_i\sim 1.40~{\rm GeV}^{-1}~{\rm for\,\, RHIC}~~~,~~~
t_i\sim 0.62~{\rm GeV}^{-1}~{\rm for\,\, LHC}
\label{FORM_TIMES}
\end{equation}
Next, we suppose that a transport theory based treatment 
of gluon scattering is applicable when the gluon
occupation number is less than unity. 
Letting $d\eta$ be  $d\eta= dz/t_i = 1$ or  $dz = t_i$, 
and using $\frac{dN}{d^3xd^3k}$ from \cite{Mueller2} at $z=0$, we find 
\begin{equation}
t_0 = \frac{c}{\alpha_S N_c}t_i.
\label{init}
\end{equation}
Here $t_0 > t_i$ is the time where the gluon occupation number is 
dilute enough that their subsequent interactions 
can be described as the small angle scattering of on-shell gluons.

\section{The Landau transport equation}

At times $t>t_0$, where $t_0$ was defined in Eq.~\ref{init}, 
a  Boltzmann--like transport equation is appropriate for 
describing the late--time evolution of the highly anisotropic 
initial gluon distribution discussed 
in the previous section. We will show below that, for central high energy 
heavy ion collisions, assuming boost invariance, the transport equation 
reduces to a Landau--type transport equation~\cite{LifPit}.

In the problem of interest, the typical scale of spatial variations is 
large compared to the typical scale of gluon--gluon scattering. The evolution 
of the system can therefore be described by a local 
Boltzmann--like kinetic equation for 
the single particle gluon distribution $f(x,p,t)$, which gives the 
density of excitations of momentum $\vec{p}$ at a point $(\vec{x},t)$. 
The Boltzmann equation is 
\be
\frac{\dd f}{\dd t} + \vec{v}\cdot\vec{\nabla}f = C(f)\,,
\label{Boltzmann1}
\ee
where $C(f)$ is the collision integral which represents the change, 
due to collisions, per unit time,
in the number of particles per unit phase space volume.

Let us first consider the left hand side of the above equation. We will 
assume that the transverse dimensions of the collision volume in 
central heavy 
ion collisions are sufficiently large that the initial expansion of the 
system is one--dimensional. The distribution $f$ then only depends on 
$z$, the co--ordinate corresponding to the collision axis. Since it is 
independent of the transverse co--ordinate $x_t$, Eq.~\ref{Boltzmann1} 
has the form
\be
\frac{\dd f(\vec{p},z,t)}{\dd t} + v_{p_z}\frac{\dd f(\vec{p},z,t)}{dz} = 
C\left[f(\vec{p},z,t)\right]\, ,
\label{Boltzmann2}
\ee
where $v_{p_z}= p_z/|\vec{p}|$. A further simplification follows 
from the assumption that the central rapidity 
region is approximately Lorentz invariant under boosts~\cite{Bjorken}. 
As was pointed out by Baym many years ago~\cite{Baym}, this assumption 
greatly simplifies the problem of solving Eq.~\ref{Boltzmann2} since it
relates the distribution function at different $z$'s in the central region. 
It therefore 
suffices to compute $f$ in the zero--rapidity slice alone. Because $f$ is 
a scalar under Lorentz transformations, it satisfies the 
relation
\be
f(p_t,p_z,z,t) = f(p_t,p_z^\prime,t) \, .
\label{fscalar}
\ee
Here $p_z^\prime = \gamma (p_z-up)$, the transformation velocity 
$u=z/t$, and $\gamma = (1-u)^{-1/2}=t/\tau$, where the proper time 
$\tau=\sqrt{t^2-z^2}$.

Computing ${\dd f}/{\dd z}$ using 
Eq.~\ref{Boltzmann2}, the Lorentz transformation relation, 
${\dd \tau}/{\dd z}|_{z=0}=0$, and ${\dd p_z^\prime}/{\dd z}|_{z=0} = -p/t$, 
one finds~\cite{Baym} at $z=0$, 
\be
v_{p_z}\frac{\dd f}{\dd z} = -\frac{p_z}{t}\,\frac{\dd f}{\dd p_z} \, .
\ee
The Boltzmann equation now reduces to
\be
\frac{\dd f(p_t,p_z,t)}{\dd t}|_{p_z t} \equiv \left(\frac{\dd}{\dd t} 
-\frac{p_z}{t}\,\frac{\dd}{\dd p_z}\right)f(p_t,p_z,t)=C\left[f(p_t,p_z,t)\right] \, .
\label{Boltzmann3}
\ee
Note that, as a consequence of our manipulations, assuming uniform
distributions in the transverse direction and boost invariance in the
longitudinal direction, the single particle distribution $f(x,p,t)$ is
now expressed as a function of $\vec{p}$ and $t$ alone.
The kinetic equation above, in the absence of collisions, has the free
streaming solution for $f$ \cite{Baym}, namely, 
\begin{equation}
f(t,p) = \exp{\left(-\beta_0\sqrt{p_{\bot}^2+p_z^2t^2/\gamma_0^2}\right)}\, ,
\label{FREE_STREAM}
\end{equation}
where $\beta_0$ and $\gamma_0$ are constants.

We shall now consider the collision term $C[f]$ on the right hand side 
of the above equation. 
If we assume that the changes in momentum $q$ in a collision are
small, namely, $q/Q_s\ll 1$,
we can treat the collision integral as simple diffusion in
momentum space. 
Following the discussion in Lifshitz and Pitaevskii~\cite{LifPit}, one 
can write 
\be
C[f] = -\frac{\dd s_{\alpha}}{\dd p_{\alpha}}\, ,
\ee
where the flux 
\begin{equation}
s_{\alpha} =   
\left(\frac{\alpha N_c}{\pi}\right)^2\,
{\cal L}\int d^3\pp\,\left(
f\frac{\partial f^{\prime}}{\partial\ppr_{\beta}} -
f^{\prime}\frac{\partial f}{\partial\pp_{\beta}}
\right)\left(
\delta_{\alpha\beta}(1-\vec{v}\cdot\vec{v^{\prime}})+v_{\alpha}v^{\prime}_{\beta}
+v_{\beta}v^{\prime}_{\alpha}\right)\, .
\label{SS}
\end{equation}
Note that $\vec{v}=\vec{p}/|\vec{p}|$.

We have used in Eq.~\ref{SS} 
the cross--section for gluon--gluon elastic scattering, 
which in the limit of small momentum transfer squared $q^2=-{\hat t}$, is
\begin{equation}
\frac{d\sigma}{d\hat{t}} = -\left(\frac{\alpha_S N_c}{\pi}\right)^2\frac{(2\pi)^3}{2(N_c^2-1)}\,\frac{1}{\hat{t}^2}\, .
\end{equation}
The collision integral contains a logarithmic collinear divergence arising 
from small angle scattering. It is represented here by ${\cal L}$, 
defined as 
\begin{equation}
{\cal L} = \int^{q_{\rm max}}_{q_{\rm min}}\frac{dq}{q} \, .
\label{LL}
\end{equation} 
The minimal momentum transfer is, $q_{\rm min} = Q_s\theta_{\rm min}$,
where $\theta_{\rm min}$ is the minimum scattering angle. Also,
$q_{\rm max}$ is the maximum momentum transfer. How one may determine 
$q_{\rm min}$ and $q_{\rm max}$ is discussed in the next section.

Having started from Eq.~\ref{Boltzmann1}, combining Eq.~\ref{Boltzmann3} and 
Eq.~\ref{SS}, we arrive at the considerably simpler expression, the 
Landau equation~\cite{LifPit,ChapCowl}
\begin{equation}
\frac{\dd f}{\dd t}-\frac{p_z}{t}\frac{\dd f}{\dd p_z} =
\lambda\, n\, {\cal L}\,\nabla^2_{\vec{p}}f+2\lambda\,
n_{-1}\,{\cal L}\,\nabla_{\vec{p}}\cdot (\vec{v}f),
\label{Landau_1}
\end{equation}
with the definitions 
\begin{equation}
n(t) = g_G\int \frac{d^3p}{(2\pi )^3}\,f(t,\pp)~~~,~~~
n_{-1}(t) = g_G\int \frac{d^3p}{(2\pi )^3}\,\frac{f(t,\pp)}{|\pp|}~~~,~~~
\lambda = 2\pi\alpha_S^2\frac{N_c^2}{N_c^2-1}\, .
\label{LAMBDA}
\end{equation}
Here, $g_G=2(N_c^2-1)$ and $\nabla_{\vec{p}}$ refers to differentiation
with respect to $\vec{p}$.

Multiplying both sides of Eq.~\ref{Landau_1} by $d^3p$ and integrating, one
finds
\be
\frac{d}{dt}(tn) = 0\, ,
\ee
namely, the number density of gluons has the behavior $n \propto 1/t$.  
Taking the second moment of Eq.~\ref{Landau_1}, one finds the following exact 
relation in the central region~\cite{Baym}
\be
\frac{\dd \epsilon}{\dd t}+ {\left(\epsilon +P_L\right)\over t} =0 \, ,
\label{ehydro}
\ee
where the local energy density is given by
\be
\epsilon(t) = g_G\int \frac{d^3p}{(2\pi)^3}\,p\,f(p,t) \, ,
\label{ehydro1}
\ee
and the longitudinal pressure is
\be
P_L(t) =g_G\int \frac{d^3p}{(2\pi)^3}\, {p_z^2\over p}\, f(p,t) \, .
\label{ehydro2}
\ee
In the hydrodynamic regime where $P_L=\epsilon/3$, the ideal gas
equation of state gives $\epsilon\sim t^{-4/3}$.  We will return to a
discussion of these quantities later on in the paper.

Following Mueller, it is convenient to define the quantities $\eta =
tn$ and $\eta_{-1} = tn_{-1}$.  The Landau equation can thus be
re--written as
\begin{equation}
t\frac{\dd f(t,\pp )}{\dd t}-p_z\frac{\dd f(t,\pp )}{\dd p_z} =
\lambda \eta \,{\cal L}\,\nabla^2_{\vec{p}}f(t,\pp )+2\lambda
\eta_{-1}(t)\,{\cal L}\,\nabla_{\vec{p}}\cdot (\vec{v}f(t,\pp )).
\label{BE_FINAL}
\end{equation}
If $\eta_{-1}$ is taken to be independent of time, this equation 
is the Fokker--Planck equation: the coefficient of the 
first term would  be the diffusion coefficient, while that of 
the second would be the coefficient of friction.

\section{Screening of collinear divergences in the Landau transport 
equation}

In the one gluon exchange approximation, as in Coulomb scattering, the 
transport integral in Eq.~\ref{SS} has a logarithmic 
collinear divergence--represented by ${\cal L}$.  In Ref.~\cite{Mueller1}, 
the condition applied at early times was that in the maximal distance 
$1/q_{\rm min}$ corresponding to the minimal momentum transfer $q_{\rm min}$, 
one has at most one scattering~\footnote{
This condition satisfies the requirement 
that the soft field (due to all the other hard gluons) that a particular hard 
gluon scatters off be at most of size $1/g$. 
This is so because the freed gluons come from the saturated component of the 
nuclear wavefunction of size $1/g$--see the discussion in section 2.}.
From this condition, by using the initial gluon distribution, Mueller finds at 
the earliest times that 
\be
M_m^2 = \left({c\alpha_S N_c\over \pi}\right)^{2/3}\,{Q_s^2\over {(Q_s t)^{2/3}}} \, .
\label{massM}
\ee
A more general expression, 
for any arbitrary number density , is easily derived to be
\begin{equation}
M_m^2 = \left( \left(\frac{\alpha_SN_c}{\pi}\right)^2\frac{(2\pi)^3}{2(N_c^2-1)}n(t)\right)^{2/3}.
\label{MASS_M}
\end{equation}
\begin{figure}[tbh]
\centering \leavevmode
%\epsfxsize=4.0in
%\epsfysize=7.0in
%\epsfbox{beamline.ps}
\psfig{file=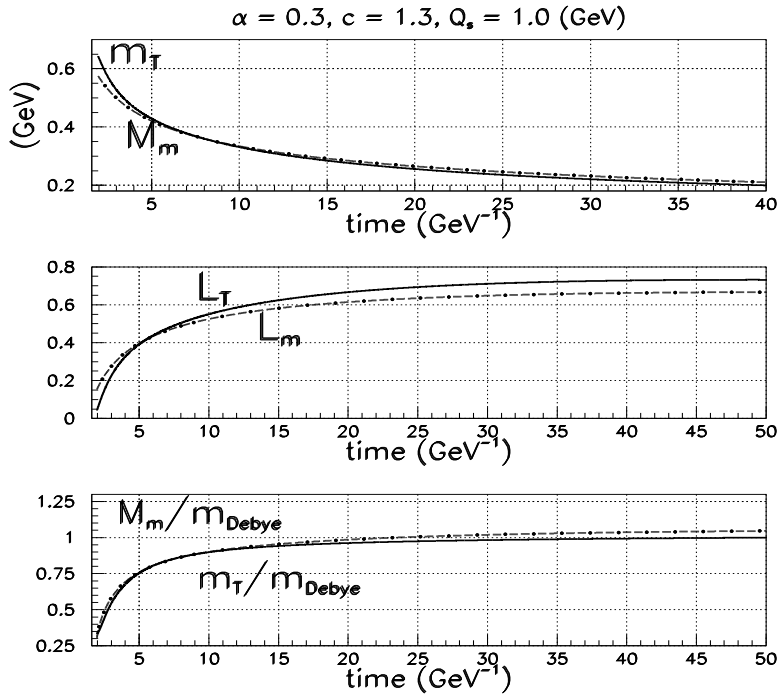,height=4.5in,width=4.in}
\caption{The dynamical 
screening masses, $M_m$ from eq. (\ref{MASS_M}) obtained from
\cite{Mueller1} and the $m_{\rm T}$ from eq. (\ref{Debye1}) obtained from 
linear
response theory \cite{BiroMuller} plotted as functions of time in GeV$^{-1}$ 
units. Also plotted versus time 
are the logarithms ${\cal L}_m$ and  ${\cal L}_t$
obtained by combining Eq.'s  (\ref{MASS_M}) and (\ref{Debye1}) respectively 
with Eq. (\ref{LL1}).}
\label{FIG_MASS}
\end{figure}

Alternately, one can use the more conventional prescription of 
regulating the logarithmic collinear divergence in ${\cal L}$ through 
the exchange of dressed gluons characterized by a 
Debye screening mass. A kinetic theory expression for the 
Debye screening mass can be derived from the screening
of a time-like gluon propagator in a medium of gluon 
excitations~\cite{BiroMuller}. 
One obtains 
\begin{equation}
m_D^2 = -\frac{\alpha_S N_c}{\pi^2}\lim_{|\vec{q}|\rightarrow 0}
\int d^3p\,\frac{|\vec{p}|}{\vec{q}\cdot\vec{p}}\,\vec{q}\cdot\vec{\nabla}_{\vec{p}}\,f(t,\vec{p}).
\label{Debye}
\end{equation}
One can check that for an equilibrium Bose distribution, one recovers
the standard result: $m_D^2=4\pi\alpha_S N_c T^2/3$.  
For our initial conditions, it is the screening of the longitudinal 
gluons in the transverse direction that is relevant. Performing an 
integration by parts, this transverse Debye mass is given by the 
expression~\cite{DumitruNayak}
\be
m_T^2 = \frac{\alpha_S N_c}{\pi^2}\,\int {d^3 p\over |\vec{p}|}\, f(p) \, .
\label{Debye1}
\ee

At the very early stage of the $2\rightarrow 2$ scattering, the
expression in Eq.~\ref{Debye1} is, in principle, 
less reliable since the system is
completely out of equilibrium. However, at later times, as the system
approaches equilibrium, $m_D$ should be more reliable. 
One could attempt to parameterize the infrared
cut-off of small angle scattering such that it interpolates between
the two limits. However, at present there is no theoretical
justification of any particular form. Since collisions become more
frequent as one approaches equilibrium, it is likely that the Debye
mass is a more reliable cut-off to use. At any rate, it is important 
that one at least have the right limit as the system approaches equilibrium.

In Refs.~\cite{Mueller1,Mueller2}, the value of $q_{\rm max}$ is held fixed at 
early times. This is because if one re--writes ${\cal L}$ in Eq.~\ref{LL} as
\be
{\cal L} =\ln\left(\frac{\langle p_t\rangle}{m_T}\right) \, ,
\label{LL1}
\ee
it is argued, on 
the basis of analytical approximations to Eq.~\ref{PT} below,  
that $\frac{\langle p_t\rangle}{Q_s}$ changes appreciably only over a long 
time scale ($t\sim e^{\frac{1}{\sqrt{\alpha_S}}}$). 
Since the Debye mass has a stronger 
time dependence, ${\cal L}$ grows with time. On account of the approximations 
made in arriving at this conclusion, it is useful to check whether this 
behavior of ${\cal L}$ is obtained by computing $\langle p_t\rangle$ 
and $m_T$ self--consistently at each step in our numerical simulations~\footnote{We thank A. Dumitru for discussions on this point.}. We indeed do find, 
as shown in  
in Fig.~\ref{FIG_MASS}, that ${\cal L}$ grows rapidly initially as a function of time 
but levels out to a constant as the system approaches equilibrium .

In Fig.~\ref{FIG_MASS}, we plot the infrared masses in Eqs.~\ref{MASS_M} and
~\ref{Debye1}, as well as the corresponding contribution ${\cal L}$ to
the collision integral, as a function of time (in units of
$1/Q_s$). One sees that while the relative difference at early times
is large, it is much less so at later times. We have also solved the
transport equation using both cut-offs. While there is 
some quantitative difference between the two, they do not affect our
qualitative conclusions. For the rest of this work, we will therefore
use the Debye mass in Eq.~\ref{Debye1}.

%\begin{figure}[htb]
%\centering \leavevmode
%\psfig{file=screen.ps,height=5in,width=4in}
%\caption{ The dynamical screening mass $m^2$ from
%(\ref{Debye}) and $L$ from (\ref{L_MASS})
%potted as functions of time. $N_c=3$, $c=1$, $\alpha = 0.3$ and $Q_s=1$.}
%\label{SCR_3}
%\end{figure}

Debye screening occurs due to the exchange of longitudinal gluons. One
may worry whether the screening of transverse gluons will introduce an
additional scale since the static magnetic mass is parametrically
larger than the Debye mass. However, the conventional wisdom is that,
for transport cross--sections, magnetic screening is also cut off in
the infrared by the Debye mass~\cite{Baymetal,LeBellac}.  The same,
for instance, is not true of color transport--the color conductivity
is regulated by a magnetic mass of order $g^2
T$~\cite{SGHASYB}. The dynamical screening of infrared
divergences in $2\rightarrow n$ ($n\geq 2$) transport cross-sections
merits further study.

\section{Results from numerical solution of the Fokker-Planck Equation}

We will first consider, as a ``warmup exercise'',  
the solution of Eq.~\ref{BE_FINAL} for the case
where $\eta_{-1}$ is held constant. The Landau equation then reduces
to the Fokker--Planck equation.  To check the
consistency of our numerical simulation, we will compare, in detail, the
numerical solution of the Fokker--Planck equation to the early time
analytical solution found by Mueller in Refs.~\cite{Mueller1} and
~\cite{Mueller2}. We will discuss solutions to the non--linear 
Landau equations in the next section.

Following Mueller, we
first define $\tilde{f} = tf$, ${\cal L} =
\frac{1}{3}\log(t/t_0)\equiv \frac{1}{3}\xi$, where $t_0$ is a
constant defined in Eq.~\ref{init}. (As discussed in the previous section, 
this assumption is justified at early times.) Scaling $\lambda/3\rightarrow
\lambda$, we can rewrite (\ref{BE_FINAL}) as
\begin{equation}
\left(\frac{\partial}{\partial\xi}-\frac{\partial}{\partial p_z}p_z
\right)\tilde{f} =
\lambda\,\eta\,\xi\,\nabla^2_{\vec{p}}\tilde{f} +
2\lambda\,\eta_{-1}\,\nabla_{\vec{p}}\cdot(\vec{v}\tilde{f}).
\label{FP}
\end{equation}
Eq. (\ref{FP}) is a Fokker--Planck equation if $\eta_{-1}$ is held
fixed.  If one integrates Eq.~\ref{FP} with respect to $d^3p$,
$d^3p\,p_{\bot}^2$ and $d^3p\,p_z^2$ we obtain Eqs. (17), (18)
and (19) in Ref.~\cite{Mueller2},
\begin{equation}
\frac{d}{d\xi}\eta = 0 \, ,
\label{17}
\end{equation}

\begin{equation}
\frac{d}{d\xi}\langle p^2_z\rangle  + 2\langle p_z^2\rangle  =
2\lambda\,\eta\,\xi-
\frac{\lambda\eta_{-1}\xi}{\eta}\tilde{P}_L(\xi)\, ,
\label{PZ}
\end{equation}
and

\begin{equation}
\frac{d}{d\xi}\langle p^2_{\bot}\rangle  = 
4\lambda\eta\xi\left(1-\frac{\eta_{-1}\eta_{+1}}{\eta^2}\right)+
\frac{\lambda\eta_{-1}\xi}{\eta} \tilde{P}_L(\xi)\, ,
\label{PT}
\end{equation}
where $\tilde{P}_L=tP_L$ is defined by eq. (\ref{ehydro2}).
These equations are exact.
We can therefore perform the following consistency check on our solution.
With the initial conditions in Eq.~\ref{idis} 
and the boundary condition that
$f(p_x,p_y,p_z,t)=0$ if any $p_x$, $p_y$ or $p_z \rightarrow\infty$, we
numerically determine $f$ for all time, directly compute $\langle
p_{\bot}^2\rangle$  and $\langle p_y^2\rangle$, and check to see if
Eqs. (\ref{17}) to (\ref{PT}) are satisfied.
Mueller supposes that at early times the first term on the l.h.s and
the last term on the r.h.s of Eq.~\ref{PZ} are small, thereby
 yielding the solution
\begin{equation}
\langle p_z^2\rangle  = \lambda\eta\xi.
\label{M_PZ}
\end{equation}

\begin{figure}[tbh]
\centering \leavevmode
\psfig{file=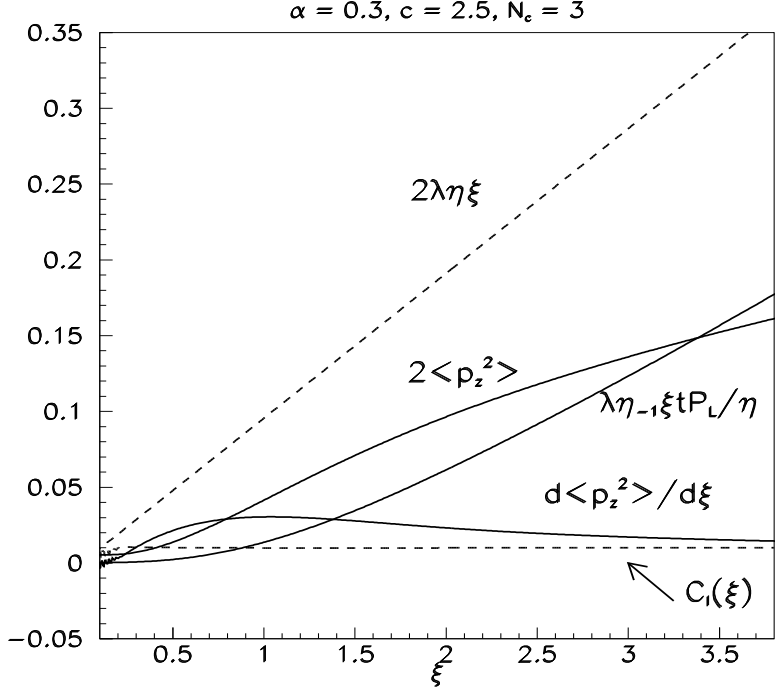,height=4.5in,width=4in}
\caption{Numerical solution to the Boltzmann equation with constant
$\eta_{-1}$  where each term in Eq. (\ref{C1}) is plotted as a function 
of $\xi =\ln(t/t_0)$.}
\label{FIG1}
\end{figure}

It is demonstrated in Fig.~\ref{FIG1} that this approximation, in
practice, is not a very good one. At very early times, if $\xi \sim 0$
we see that the first term on the l.h.s of Eq.~\ref{PZ} is of the same
order as the remaining terms. We cannot therefore ignore it.  We can
ignore the term proportional to $\tilde{P}_L(t)$ in Eq.~(\ref{PZ}) for
early times. Then Eq.~(\ref{PZ}) becomes:
\[
\frac{d}{d\xi}\langle p^2_z\rangle  + 2\langle p_z^2\rangle  =
2\lambda\eta\xi \, .
\]

Its solution is 

\begin{equation}
\langle p_z^2\rangle  = \lambda\eta\xi+\frac{\lambda\eta}{2}(\exp{(-2\xi)}-1).
\label{ME_PZ}
\end{equation}

This solution (see Fig. \ref{FIG3}) agrees much better with 
the numerical results than Eq. (\ref{M_PZ}) does. We define $C_I(\xi)$ as

\begin{equation}
2\lambda\eta\xi-\frac{\lambda\eta_{-1}\xi}{\eta}\tilde{P}_L(\xi)-
\frac{d}{d\xi}\langle p^2_z\rangle  - 2\langle p_z^2\rangle  = C_I(\xi),
\label{C1}
\end{equation}

where $C_I(\xi)$ should be small if Eq. (\ref{PZ}) is
satisfied. Fig. \ref{FIG1} verifies indeed that this is so.

Next, look at Eq.~(\ref{PT}). If we assume that both $\eta_{-1}$ and
$\eta_{+1}$ are constant with respect to time and determined by
the initial condition Eq. (\ref{idis}),

\begin{equation}
\eta = c\frac{N_c^2-1}{4\pi^2\alpha_S N_c}Q_s^2~~~,~~~
\eta_{-1} = c\frac{N_c^2-1}{2\pi^2\alpha_S N_c}Q_s~~~,~~~
\eta_{+1} = c\frac{N_c^2-1}{6\pi^2\alpha_S N_c}Q_s^3.
\label{ETA_BC}
\end{equation}

Using Eq.~(\ref{ETA_BC}) we re--write Eq.~(\ref{PT}) as:

\begin{equation}
\frac{d}{d\xi}\langle p^2_{\bot}\rangle =
-\frac{4}{3}\lambda\eta\xi+\frac{\lambda\eta_{-1}\xi}{\eta}\tilde{P}_L(\xi).
\label{ME18}
\end{equation}

Again the term containing $\tilde{P}_L$ is small. We therefore solve the
equation above, with the appropriate boundary conditions, to find

\begin{equation}
\langle p^2_{\bot}\rangle  = \frac{Q_s^2}{2}-\frac{2}{3}\lambda\eta\xi^2.
\label{PT_ME}
\end{equation}

As previously, define $C_{II}(\xi)$ such that

\begin{equation}
\frac{d}{d\xi}\langle p^2_{\bot}\rangle +
\frac{4}{3}\lambda\eta\xi-\frac{\lambda\eta_{-1}\xi}{\eta} \tilde{P}_L(\xi)
=C_{II}(\xi ).
\label{C2}
\end{equation}

\begin{figure}[tbh]
\centering \leavevmode
\psfig{file=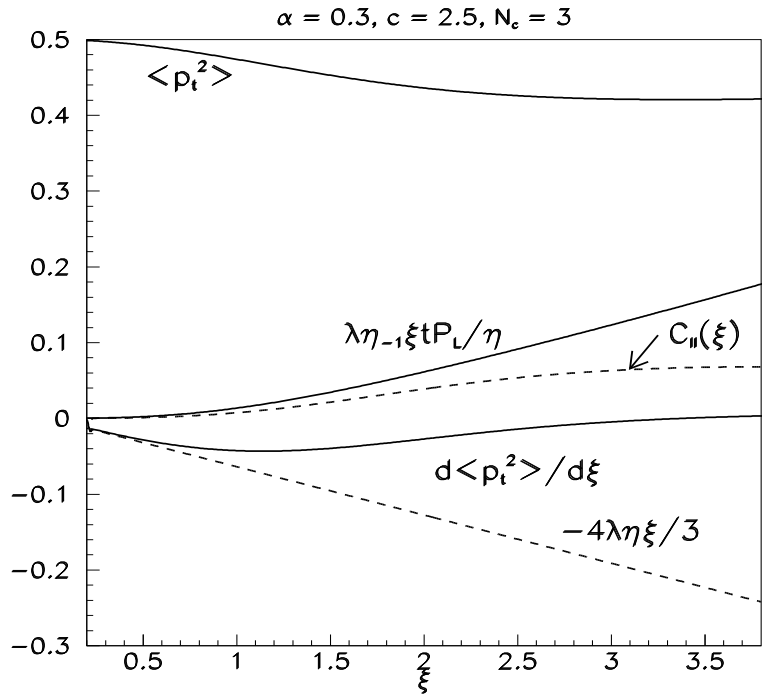,height=4.5in,width=4.in}
\caption{Each term in Eq. (\ref{C2}) plotted as a function of 
$\xi =\ln(t/t_0)$.}
\label{FIG2}
\end{figure}

Fig.~\ref{FIG2} verifies that Eq. (\ref{C2}) is satisfied for early times.

Finally, we check to see how well Eq. (\ref{17}) is satisfied
numerically. Fig. \ref{FIG3} plots the numerical result for $\eta$
vs. $t$ and demonstrates that $\eta$ is constant with respect to $t$.
As discussed earlier, Fig. \ref{FIG3} also compares the 
numerical result for $\langle
p_z^2\rangle $ and $\langle p_{\bot}^2\rangle $ with the analytical 
solutions of Eqs. (\ref{ME18}) and (\ref{PZ}) given by Eqs. (\ref{ME_PZ}) and
(\ref{PT_ME}). 

\begin{figure}[tbh]
\centering \leavevmode
%\epsfxsize=4.0in
%\epsfysize=7.0in
%\epsfbox{beamline.ps}
\psfig{file=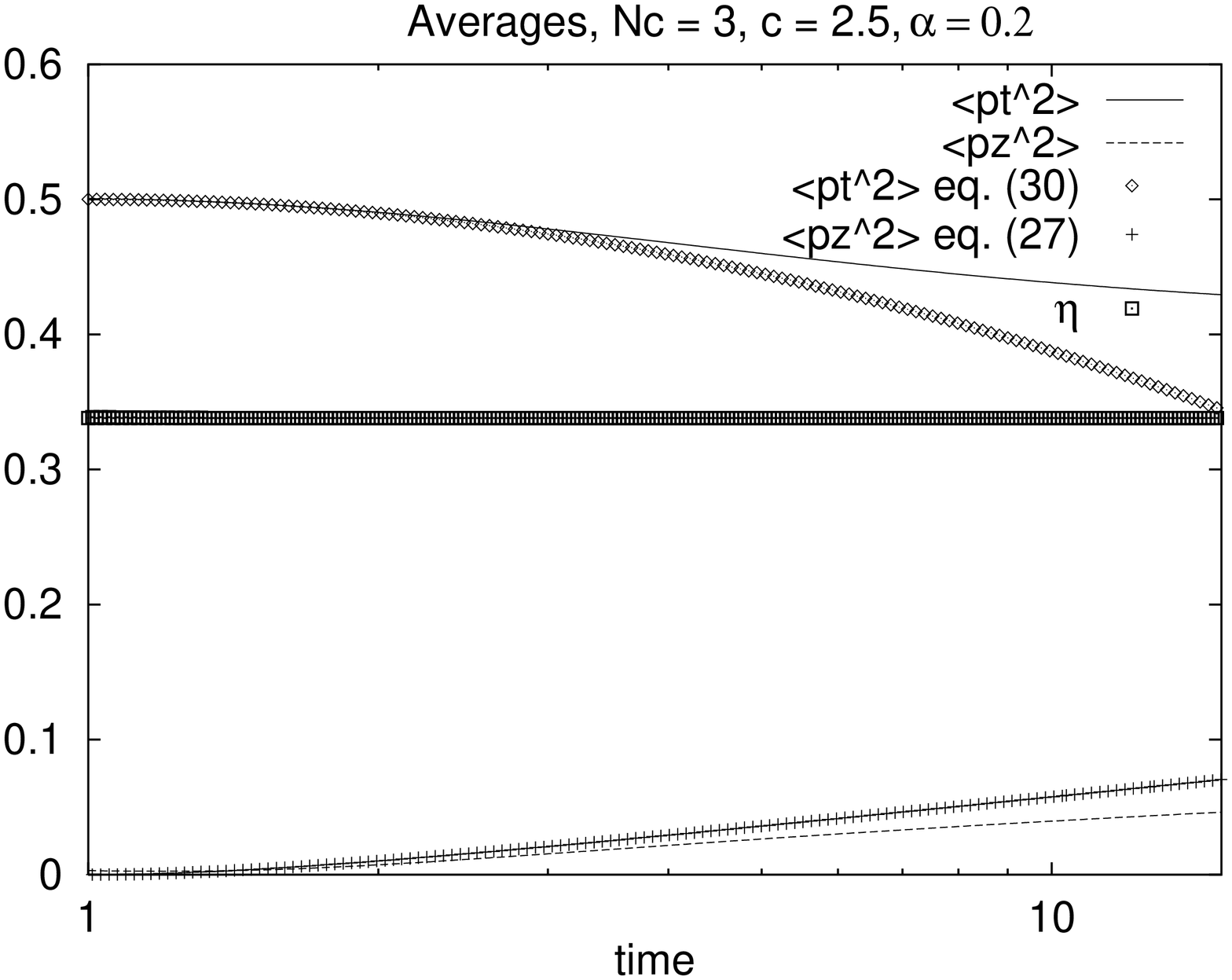,height=3.5in,width=5.5in,angle=-90}
\caption{
Numerical results for $\langle p_z^2\rangle $ and 
$\langle p_{\bot}^2\rangle$, plotted as a function of time, 
compared with the analytical approximations in 
Eqs. (\ref{ME_PZ}) and (\ref{PT_ME}) . Also plotted is
$\eta$ as a function of time.}
\label{FIG3}
\end{figure}

\section{Results from numerical solution of the Landau transport equation}

\begin{figure}[tbh]
\centering \leavevmode
\psfig{file=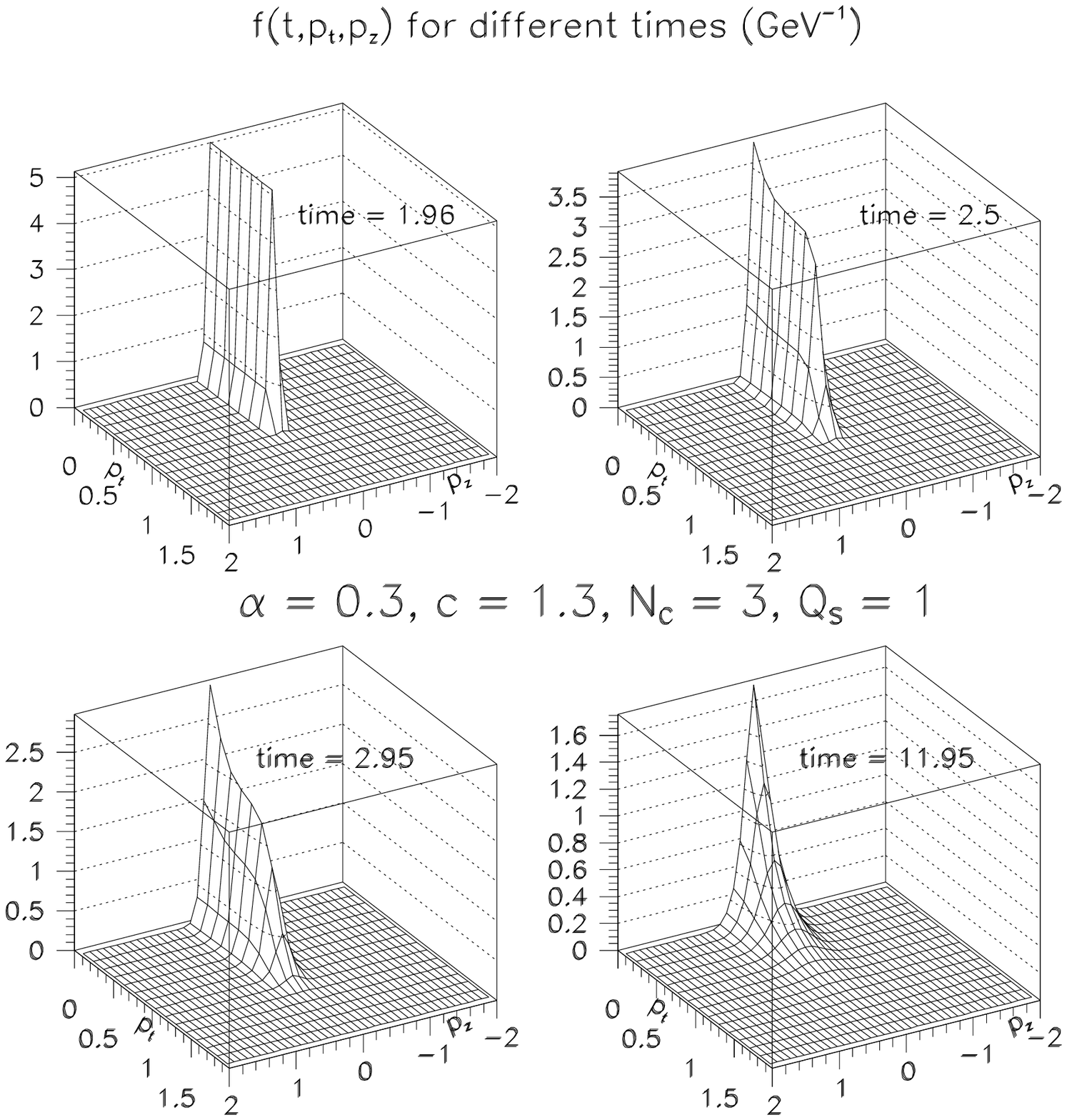,height=5.in,width=4.5in}
\caption{ The number of gluons per unit of phase space $f(t,\pp)$
plotted as a function of $p_{\bot}$ and $p_z$ for different
times. $N_c = 1$, $\alpha_S = 0.3$, $c = 1.3$ and $Q_s = 1$ GeV. 
Units of time are in ${\rm GeV}^{-1}$.}
\label{F_FULL}
\end{figure}
%\clearpage

In the previous section, we numerically solved the Boltzmann equation 
using Mueller's approximation for the minimum scattering angle,
and with $\eta_{-1}$ fixed for all time (Eq. (\ref{FP}))~\footnote{Recall that
$\eta_{-1}$ is an integral representing the ``-1'' moment of $f$.}. 
Since the previous analysis is only expected
to be valid for early times, 
one must solve the general expression (\ref{BE_FINAL})
with $\eta_{-1}$ determined self consistently.

In this section, we will discuss results from numerical solutions of
the Landau equation in Eq.~\ref{BE_FINAL} employing the initial single
particle distribution given by Eq.~\ref{idis} and the dynamical
screening mass given by Eq.~\ref{Debye1}. The equation is a second
order partial integro--differential equation.
It can be solved by combining the Alternating Direction Implicit (ADI)
method and the Crank--Nicholson differencing
scheme~\cite{Recipes,STRIKWERDA}. A detailed discussion of the
numerical procedure can be found in Appendix A.

We will first begin by briefly describing the evolution of the single
particle distribution and its moments.  Next, we will discuss various
measures of kinetic equilibration and hydrodynamic flow for a system
undergoing boost--invariant one dimensional expansion. Finally, we
will discuss our numerical results for these quantities for different
values of $Q_s$ and $\alpha_S$~\footnote{We will ignore running
coupling effects in this analysis. It is argued in
Ref.~\cite{Mueller2} that running coupling effects in the evolution
are O($\sqrt{\alpha_S}$) and therefore suppressed.}.

\subsection{Single particle distributions}

The initial distribution for the numerical solution of
Eq.~(\ref{BE_FINAL}) is given by (\ref{idis}), at the initial time,
determined from Ref.~\cite{AlexRaj2}, given by Eq.~(\ref{FORM_TIMES}).
The time evolution of the single particle distribution
$f(t,p_z,p_{\bot})$ is shown in Fig.~\ref{F_FULL} for a particular set
of initial parameters.  It begins its time evolution as a delta
function in $p_z$ (represented practically by a narrow Gaussian
distribution) and as a step function in $p_{\bot}$ --- see
Eq.~(\ref{idis}).  As time proceeds, we see the $p_{\bot}$ modes get
scattered off the transverse plane in the longitudinal direction.  The
larger $p_{\bot}$ modes decrease rapidly as the $p_z$ distribution
widens out. At about $t= 11.95~{\rm GeV}^{-1}$, the $p_z$ distribution
is at its broadest extent.

\begin{figure}[tbh]
\centering \leavevmode
\psfig{file=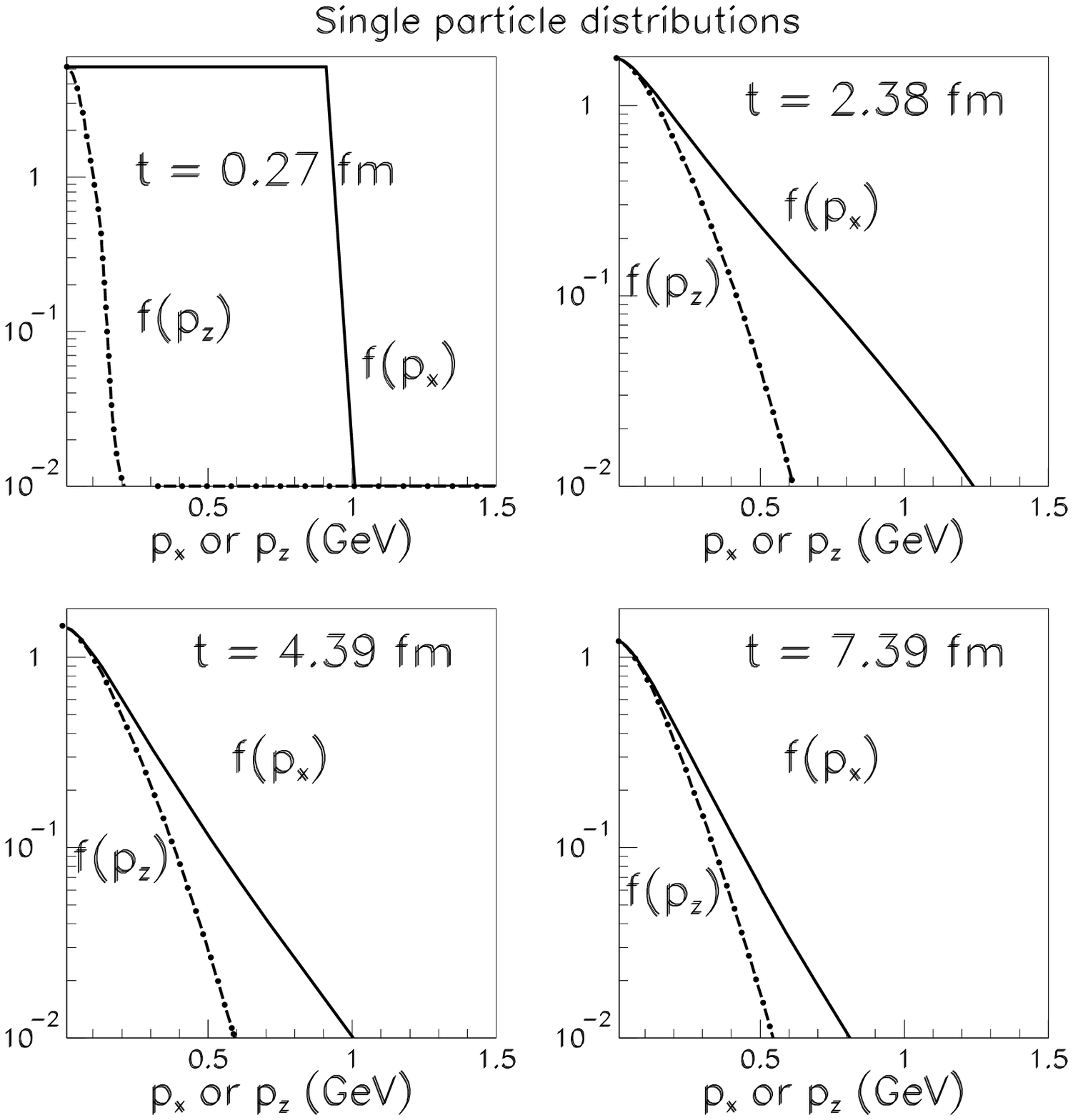,height=4.5in,width=4in}
\caption{ The single particle distributions, for fixed $p_x$ or $p_z$ plotted 
as a function of $p_z$ and $p_x$ respectively, for different times in the 
evolution of the distribution. The results are for $\alpha_S=0.3$, $Q_s=1$,  
$c=1.3$ and $N_c=3$.}
\label{SINGPART}
\end{figure}

\begin{figure}[tbh]
\centering \leavevmode
\psfig{file=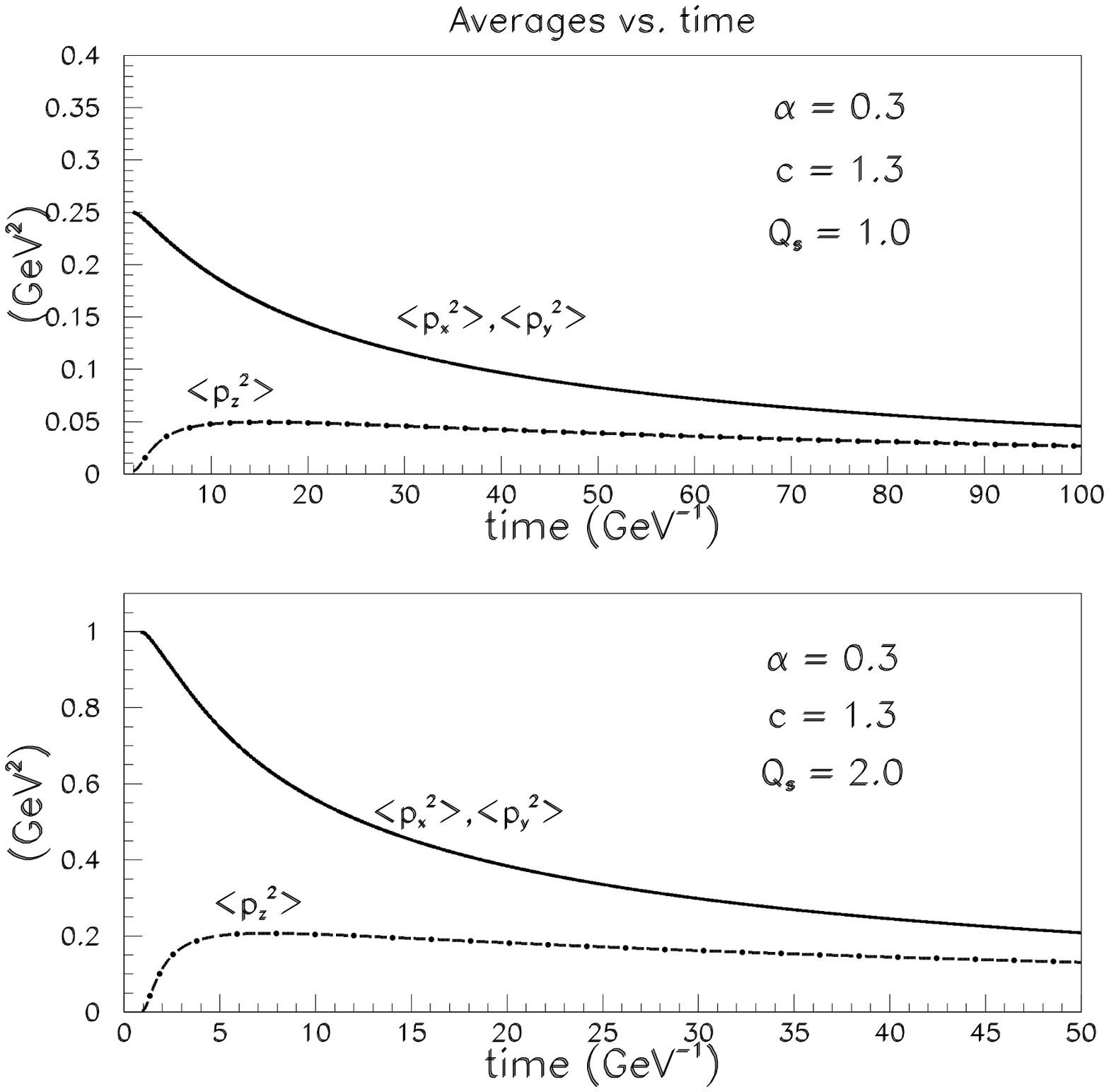,height=4.5in,width=4in}
\caption{ The averages $\langle p_x^2\rangle$, $\langle p_y^2\rangle$
and  $\langle p_z^2\rangle$ versus time in GeV$^{-1}$ units for
different values of $\alpha_S$, $c$ and $Q_s$. In all cases, $N_c=3$. }
\label{AVE_DEBYE}
\end{figure}

The behavior of the single particle distributions are seen more clearly 
in Fig.~\ref{SINGPART}, where we have plotted $f$, as a function of 
$p_z$ or $p_x$, for fixed $p_x$ and $p_z$ respectively. We see initially 
that the two distributions are widely different. At time proceeds, 
they converge relatively rapidly at soft momentum but much more slowly 
in the tails. If thermal equilibrium is defined strictly as the distributions 
being completely isotropic, this condition is reached only asymptotically, 
if at all.

Fig.~\ref{AVE_DEBYE} shows, for particular initial parameters, 
the averages $\langle p_x^2\rangle $, $\langle p_y^2\rangle $
and $\langle p_z^2\rangle $ as functions of time. $\langle
p_z^2\rangle $ starts at zero and quickly rises before converging slowly
to $\langle p_x^2\rangle $ and $\langle p_y^2\rangle $, which, in turn,  
decrease monotonically with time. The non--trivial behavior of $\langle 
p_z^2\rangle$ is because the system is undergoing longitudinal expansion. In 
a box at rest, one expects that $\langle p_z^2\rangle$ will show more of a 
monotonic behavior before leveling off. It again appears from 
Fig.~\ref{AVE_DEBYE} that the convergence to isotropic distributions 
is very slow. This is particularly so since the second moment of the 
distribution weights the high momentum tail unduly. The latter,
 as we observed in 
Fig.~\ref{SINGPART}, takes longer to equilibrate. 

Interestingly, as we will discuss in the following, the convergence of 
bulk thermodynamic observables to the expected equilibrium values is 
much more rapid. A likely explanation is that these observables are much 
less sensitive to the high momentum tail of the distribution.

\subsection{Kinetic equilibrium and hydrodynamics}

From Figs.~(\ref{SINGPART}) and (\ref{AVE_DEBYE}), it appears that the
distribution, as a whole, become isotropic only asymptotically. This
statement is particularly true of the tails, in $p_z$ and $p_t$, of
the distributions.  The distributions agree more closely at softer
momenta.  Nevertheless, as we shall discuss below, when we look at
thermodynamic signatures of equilibrium, they converge relatively
rapidly to the expected behavior in a fluid undergoing
boost--invariant 1--dimensional expansion.

Let us first discuss what this expected behavior is. Since we only 
consider $2\rightarrow 2$ processes in this work, the total number of 
gluons is fixed. The equilibrium solution 
of the classical Boltzmann equation, Eq.~(\ref{BE_FINAL}), thus has the form
\begin{equation}
f(t,p) = \exp{\left[\beta(t)(\mu(t)-p)\right]}\, .
\label{F_EQ}
\end{equation}
Here $\mu(t)$ is the chemical potential, and $\beta(t) = 1/T(t)$, where $T$ is 
is the temperature. Substituting this equilibrium distribution in 
Eq.~(\ref{LAMBDA}), one obtains 
\begin{equation}
n(t) = 2\,\frac{(N_c^2-1)}{\pi^2}\,T^3(t)\,\exp\left({\frac{\mu}{T}}\right)~~~,~~~
n_{-1}(t) = \frac{(N_c^2-1)}{\pi^2}\, T^2(t)\,\exp\left({\frac{\mu}{T}}
\right)\, .
\label{ETAS}
\end{equation}
The average energy density is defined to be 
\begin{equation}
\epsilon(t) = g_G\int \frac{d^3p}{(2\pi )^3}\,|\vec{p}\,|\,f(t,p)\, .
\label{ENERGY}
\end{equation}
From Eq.~(\ref{ENERGY}) and Eq.~(\ref{ETAS}), 
the energy per particle  at equilibrium is, 
\begin{equation}
E(t) = \epsilon (t)/n(t) = 3\,T(t) \, .
\label{TEMP}
\end{equation}
Now, the entropy density of a classical Boltzmann gas is defined as 
\begin{equation}
s(t) = -g_G\int\frac{d^3p}{(2\pi )^3}\,f\log f \,\, .
\label{S_BOSE}
\end{equation}
In equilibrium, the entropy per particle is simply,
\begin{equation}
S(t) = \frac{s(t)}{n(t)}= 3-\frac{\mu}{T}\, .
\label{S_EQUIL}
\end{equation}

We noted previously, since the number of gluons is conserved, that 
$t n(t) = constant$. Also note that, since the entropy per particle is 
constant in equilibrium, we find from the above equation that 
$\mu/T=constant$. From these constraints, the system in equilibrium must 
satisfy
\be
T^3\, t = {\rm constant} \,\,\, ,\,\,\, \epsilon\, t^{4/3}={\rm constant}
 \,\, .
\label{constraint1}
\ee
Finally, recall we had defined the longitudinal pressure in the central 
slice in Eq.~\ref{ehydro2} as 
\be
P_L(t) = g_G\,\int {d^3 p\over (2\pi)^3}\,{p_z^2\over p}\,f(p,t) \, .\nonumber
\ee
One can similarly define the transverse pressure $P_T$ to be
\be
P_T(t) = g_G\, \int {d^3 p\over (2\pi)^3}\, {p_t^2\over 2p}\, f(p,t) \, .
\ee
From Eq.~\ref{ehydro}, and the above relations, the condition for ideal 
hydrodynamics is 
\be
P_T=P_L = {1\over 3}\,\epsilon \, . 
\label{constraint2}
\ee
The 
approach to equilibrium, in the sense of ``saturating'' the 
above thermodynamic (and hydrodynamic) identities, has been studied, 
in the relaxation time approximation, by several authors~\cite{Baym,
Gavin,HX1}. This approximation is also employed in studies with mini--jet 
initial conditions~\cite{EskKajRuu,DumitruNayak}.
The collision kernel in Eq.~\ref{Boltzmann3} can be written formally 
as~\cite{ChapCowl,PPVW} as
\be
C[f(p_z,p_t,t)] = -{\left(f-f_{\rm {equil}}\right)\over \theta} \,\, ,
\ee
where $\theta$, the collision or relaxation time, is in general a function 
of time and momentum. The relaxation time approximation is one where the 
momentum dependence of $\theta$ is neglected. Baym studied equilibration in 
this approximation taking $\theta$ to be a constant. He showed that the 
thermodynamic relations in Eqs.~\ref{constraint1} and ~\ref{constraint2} were 
satisfied only asymptotically in 
time~\footnote[1]{This is indeed what we would conclude from our  
study of the single particle distributions in the previous sub--section.}. 
Nevertheless, the convergence of the system to the asymptotic value is 
rapid.  The system, while not quite in local thermodynamic equilibrium, 
is sufficiently close to it that equilibrium is a good working assumption. 
Subsequently, Gavin studied equilibration in the relaxation time approximation 
assuming $\theta$ to have the time dependence~\cite{Gavin}; 
$\theta=\alpha t$, where $\alpha$ is 
a constant. Depending on the value of $\alpha$, the system approaches the 
hydrodynamic limit quickly (smaller $\alpha$'s) or free streaming (larger 
$\alpha$'s). Heiselberg and Wang~\cite{HX1}, studied the general case, 
$\theta=t^p$. They conclude that for $p < 1$ thermalization is attained, while 
the system free streams for $p > 1$. The case $p=1$ studied by Gavin  is the 
marginal one, interpolating between the two regimes for different $\alpha$'s. 
In a subsequent paper~\cite{HX2}, Heiselberg and Wang study the dependence 
of $\theta$ in finite temperature QCD, and tentatively conclude that $p\sim 
1/3 < 1$.

\subsection{Convergence to the hydrodynamic limit}

In this work, as discussed in section 3, we go beyond the relaxation time 
approximation in treating the collision kernel. It is not necessary therefore, 
once a screening mechanism is postulated, to specify the time dependence of 
the collision time. In the following, we will discuss our 
results for the thermodynamic relations stated in the previous section.

\begin{figure}[bthp]
\centering \leavevmode
\psfig{file=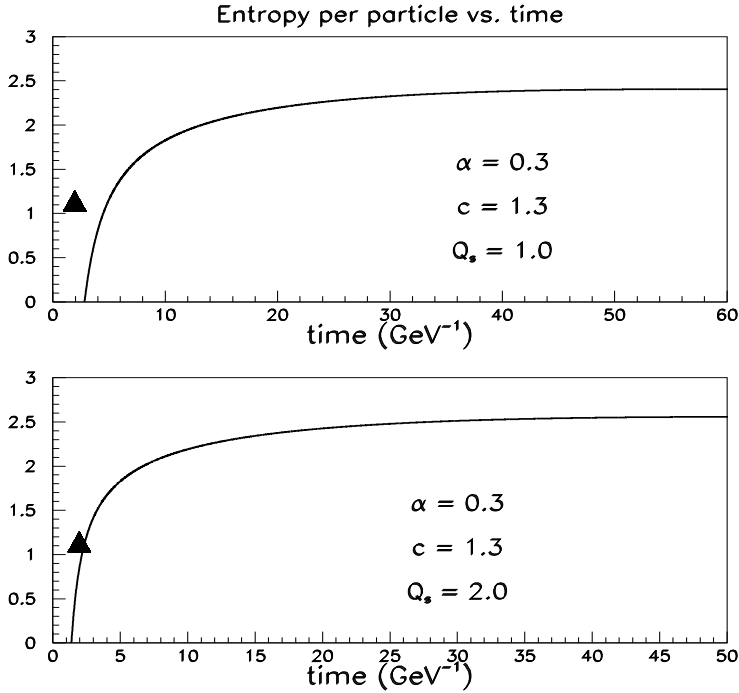,height=4.5in,width=4.in}
\caption{The entropy per particle $S$ plotted as a function of time for some
typical values of $\alpha_S$, $c$ and $Q_s$. In both cases, $N_c=3$. The
triangles denote the entropy per particle of a two dimensional Boltzmann 
gas.}
\label{ENTROPY}
\end{figure}

In Fig. \ref{ENTROPY} we show the entropy per particle, $S$, plotted
as a function of time. As shown for typical values of $\alpha_S$, $c$
and $Q_s$, it initially increases rapidly
and flattens out, monotonically approaching its asymptotic value
(thereby implying that $\mu/T$ goes to a constant as well).  The
expression in Eq.~\ref{S_BOSE}, used in the computation of
Fig.~\ref{ENTROPY}, is the correct one, except at very early times
when the well--known full expression for the entropy density of a Bose
gas should be used. Thus, where the result in Fig.~\ref{ENTROPY} is
going to zero is where the classical Boltzmann expression for the
entropy density is breaking down. We have checked that the full
expression ensures that the entropy per particle is always positive
definite. The triangles in Fig.~\ref{ENTROPY} correspond to the
entropy per particle assuming the partons are initially localized on
the two dimensional transverse plane.

\begin{figure}[tbhp]
\centering \leavevmode
\psfig{file=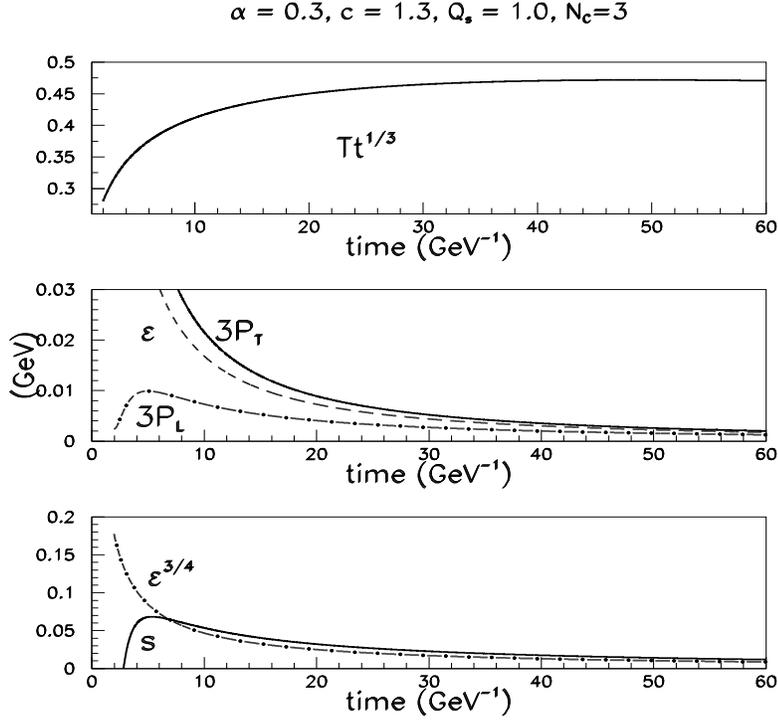,height=4.5in,width=4.in}
\caption{The top panel plots $Tt^{1/3} = \frac{\epsilon}{3n}\,t^{1/3}$
where $\epsilon$ is the energy density and $n$ the number density; the
middle panel the $\epsilon$, the longitudinal pressure $P_L$ and the
transverse pressure $P_T$; the bottom panel $\epsilon^{3/4}$ 
and the entropy density $s$; all plotted versus time in
GeV$^{-1}$ units, for typical values of $\alpha_S$, $c$, $Q_s$ and
$N_c$.}
\label{TPLOT}
\end{figure}

The convergence of other thermodynamic quantities 
to the hydrodynamic limit is shown in Fig.~\ref{TPLOT}.
Again, we note that the convergence to their asymptotic values is much 
more rapid than one would expect by looking at the single particle 
distributions alone. Our results suggest that the collision time effectively 
has a time dependence $t^p$ with $p<1$.

%\clearpage

Since the convergence to the hydrodynamic limit is only asymptotic, 
deciding when the system can be described in terms of thermodynamic 
quantities is somewhat subjective.
Here, we define the equilibrium time $t_{eq}$ as the time it takes for 
$Tt^{1/3}$ (\ref{TEMP}) and the entropy  per particle $S$
(\ref{S_EQUIL}), to reach $90\%$ of their maximum
asymptotic value (see Figs. \ref{ENTROPY} and \ref{TPLOT}). One can then 
also extract the ``initial'' temperature and chemical potential that 
correspond to $t_{eq}$ by using Eqs.~\ref{ETAS} -- \ref{S_EQUIL}.

\begin{figure}[tbhp]
\centering \leavevmode
\psfig{file=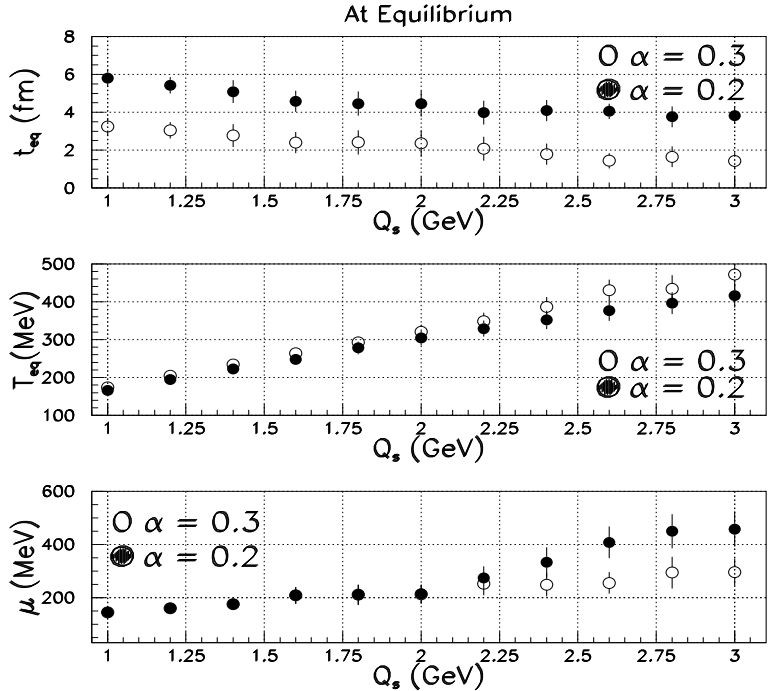,height=4.5in,width=4.in}
\caption{The estimated values of the time, $t_{eq}$, temperature
$T_{eq}$, and the chemical potential, $\mu$, at the onset of equilibrium,
plotted as functions of $Q_s$ for $\alpha_S = 0.2$ and $0.3$.}
\label{QSPLOTS}
\end{figure}

The equilibration time $t_{eq}$ is very sensitive to the values of $\alpha_S$,
$c$ and $Q_s$.  Table \ref{TAB1} shows the equilibration time
$t_{eq}$, the initial temperature $T_{eq}$, and the 
chemical potential $\mu$ for
typical values of $\alpha_S$, $c$ and $Q_s$. The parametric behavior of 
these quantities (for a fixed value of the non--perturbative constant $c$) 
as a function of $Q_s$, for two different values of the coupling constant 
$\alpha_S$, is also shown in Fig.~\ref{QSPLOTS}.
Larger values of $\alpha_S$ and $Q_s$ yield smaller $t_{eq}$.

%\clearpage

These results can be qualitatively understood as follows. One can 
show (see Eq.~\ref{PT_ME}) that the equilibration time is
parametrically 
\be
t_{eq}\sim \frac{1}{Q_s}\,\exp\left(\sqrt{\frac{2\pi}{c\alpha_S
N_c}}\right)\, . \nonumber
\ee
In Fig.~\ref{QSPLOTS}, we note that $t_{eq}$ decreases 
roughly as $1/Q_s$. Also, it is greater for smaller $\alpha_S$ as 
one would expect 
from this expression. Similarly, from requiring that $t n(t)={\rm constant}$, 
one finds for the initial temperature that  
\be
T\propto {Q_s \over \alpha_S\exp\left(\sqrt{\frac{2\pi}{c\alpha_S N_c}}
\,\right)} \, . \nonumber
\ee
We see that 
this dependence on $Q_s$ and $\alpha_S$ is confirmed in Fig.~\ref{QSPLOTS}.

\begin{table}[bth]
\begin{center}
\begin{tabular}{|c|c|c|c|c|} \hline
$\alpha_S$ & $Q_s$(GeV) &$t_{eq}$(fm) & $T_{eq}$(MeV)  & $\mu$(MeV) \\ \hline\hline
 0.1     &  1.0  & 12.24 $\pm$ 1.92 & $171.05\pm 4.9$   &$137.78\pm 19.3$ \\\hline
 0.2     &  1.0  & 5.8   $\pm$ 0.83 & $166.28\pm 5.19$  &$144.71\pm 15.22$ \\\hline
 0.2     &  2.0  & 4.48  $\pm$ 0.75 & $304.32\pm 19.39$ &$212.95\pm 31.23$ \\\hline
 0.3     &  1.0  & 3.24  $\pm$ 0.44 & $174.27\pm 5.21$  &$157.86\pm 16.18$ \\\hline
 0.3     &  1.4  & 2.77  $\pm$ 0.60 & $234.70\pm 11.60$ &$195.71\pm 31.75$ \\\hline
 0.3     &  2.0  & 2.36  $\pm$ 0.44 & $320.72\pm 22.38$ &$249.61\pm 25.06$ \\\hline
 0.3     &  2.4  & 1.80  $\pm$ 0.56 & $386.38\pm 28.48$ &$337.56\pm 26.18$ \\\hline
 0.3     &  3.0  & 1.42  $\pm$ 0.45 & $471.69\pm 36.49$ &$457.72\pm 36.17$ \\\hline
\end{tabular}
\end{center}
\caption{Equilibration times, temperatures and chemical potentials
as a function of $\alpha_S$ and $Q_s$ for $c = 1.3$ and $N_c=3$. }
\label{TAB1}
\end{table}

At RHIC, we expect (roughly) that  
$\alpha_S \sim 0.3$, $c\sim 1.3$ and $Q_s\sim~1.0$ GeV. For the idealized 
initial conditions discussed here, the corresponding 
time and temperature at kinetic equilibrium are $t_{eq}\sim
3.2~{\rm fm}$ and $T_{eq}\sim 174~{\rm MeV}$. 
At LHC, we expect $Q_s\sim~2$ -- $3$ GeV, therefore  $t_{eq}\sim
2.4~{\rm fm}$ -- $1.4~{\rm fm}$ and 
$T_{eq}\sim 321~ {\rm MeV}$ -- $472~{\rm MeV}$. Are these numbers 
realistic? Not so (especially at RHIC), 
since it is unlikely that, at these temperatures, the system is a weakly 
coupled gluonic gas. At the time scales and temperatures corresponding to  
a rapid convergence of the system to the hydrodynamic limit,  
other (likely non--perturbative) effects might become important. One cannot 
conclude definitively that this is the case because we have not considered 
realistic initial distributions nor have we discussed the importance of 
particle number changing processes. We will comment on this point in the 
final section below.

\section{Summary and outlook}

We have solved numerically a non--linear transport equation,
Eq.~(\ref{BE_FINAL}), which describes the evolution, after a heavy ion 
collision, of single particle gluon distributions in the central rapidity 
slice. The initial conditions for the solution of this equation 
are the highly anisotropic, idealized, initial conditions first discussed 
in the context of parton transport by Mueller~\cite{Mueller1,Mueller2}. 
The distributions are controlled by 
a single scale--the saturation scale $Q_s$ of parton distributions in 
the nuclear wavefunction {\it before the collision}.
Only elastic gluon--gluon scatterings 
are treated. Equilibration, in this approach, is dominated by small angle 
scattering. The collinear divergences that occur are regulated dynamically 
by a cut-off of the Debye form. We have checked that, at early times, the 
linearized Fokker--Planck equation reproduces the analytical results of 
Mueller. These analytical approximations however cannot be carried through 
to times relevant for equilibration.

We find that the tails of the initially highly anisotropic initial conditions 
converge very slowly to the expected isotropic equilibrium distribution. 
This behavior is confirmed by the behavior of $\langle p_z^2\rangle$ and 
$\langle p_x^2\rangle$. Despite the slow convergence of the single particle 
distributions to the isotropic thermal shape, thermodynamic observables 
such as the entropy per particle, the energy per particle, and the transverse 
and longitudinal pressures converge more rapidly to the hydrodynamic 
behavior expected of 1--dimensional, boost invariant expansion. The more 
rapid convergence for these observables occurs because they are more sensitive 
to softer momentum modes and less so to the high momentum tails.

Using a particular criterion for equilibration (that thermodynamic observables 
have reached $90\%$ of their asymptotic value), we extracted the equilibration 
times and the initial temperatures and chemical potentials. We have studied 
how they behave for varying values of $Q_s$ and $\alpha_S$. Even though 
small angle scattering is very inefficient (the equilibration time is 
an order of magnitude greater than the formation time), it is nevertheless 
smaller than the hydrodynamic time scale $t_{hydro}\sim R/c_0$. Here $R$ is 
the radius of the nucleus and $c_0$ is the speed of sound in the fluid. 
The relatively long equilibration times correspond to relatively low 
temperatures, $T\ll Q_s$. At these temperatures, it is unlikely that 
the system can be described as a weakly coupled gluon gas. 
It is therefore reasonable to ask whether 
other effects, be they perturbative or non--perturbative in nature, 
may significantly alter our results~\footnote[2]{For an interesting recent 
take on this topic see Ref.~\cite{AdrianMiklos}.}. We will enumerate below, in 
order of increasing complexity, those effects that are amenable to a 
weak coupling treatment.

Firstly, recall that our results were obtained for idealized initial 
conditions. More realistic initial distributions have recently become 
available~\cite{AlexRaj3}, and they are qualitatively different from the 
idealized distributions. Secondly, since the occupation numbers of the 
gluons are large initially, the effects of Bose enhancements in the 
Boltzmann equation should be taken into account. Both of these effects are
straightforward to incorporate in our approach, and results with these 
improvements will be reported shortly~\cite{JeffRaj}. 

Thirdly, as we discussed in the introduction, number changing
$2\rightarrow 3$ processes may be very important in establishing
kinetic as well as chemical
equilibrium~\cite{BiroDoorn}.
Even though these processes are suppressed by an additional power of
$\alpha_S$, they are more efficient in re--distributing the momenta.
Unfortunately, past treatments have been handicapped by uncertainties
in how one treats infrared divergences. Work in progress suggests that
this problem may be cured, and that equilibration is much more
rapid~\cite{AMDScomm}. If this is indeed the case, then initial
temperatures are relatively close to the initial saturation scale, and
the weak coupling treatment of equilibration will have been
self--consistent.  

Ultimately, the goal is relate the rich variety of hadronic and
electromagnetic spectra that will soon be available at RHIC (and some
years later, at LHC) to properties of the initial nuclear
wavefunction, that may independently be probed in deeply inelastic eA
or in pA collisions. This paper is a first quantitative step in that
direction.

\section{Acknowledgments}
We thank Jean-Paul Blaizot, Edmund Iancu, Keijo Kajantie, 
Alex Krasnitz, Dirk
Rischke, Edward Shuryak, and Xin-Nian Wang for useful comments. 
In particular, we would like to thank
Adrian Dumitru and Miklos Gyulassy on the one hand, and Al Mueller and
Dam Son on the other, for illuminating discussions of their respective
works in progress.  Larry McLerran has provided wise comments and
encouragement throughout the course of this work. This work was
supported under DOE Contract No. DE--AC02--98CH10886 at BNL and 
DE-FG02-87ER40328 at the University of Minnesota.

\appendix
\section{Numerical analysis of second order partial integro-differential equations}

Eq. (\ref{BE_FINAL}) is a second order partial integro-differential equation
whose numerical solution requires some care to ensure stability. Fortunately, standard finite 
differencing schemes can be used and numerical stability guaranteed. Eq. (\ref{BE_FINAL}),
is of the form:

\begin{equation}
\frac{\dd u}{\dd t} = A(t)\left(\frac{\dd^2u}{\dd x^2}+\frac{\dd^2u}{\dd y^2}
+\frac{\dd^2u}{\dd z^2}\right) + 
B(t)\left(\frac{\dd u}{\dd x}+\frac{\dd u}{\dd y}+\frac{\dd u}{\dd z}\right),
\label{DIF_ONE}
\end{equation}
where $u = u(t,x,y,z)$. 

To discretize Eq. (\ref{DIF_ONE})
we first define a multi-dimensional grid:

\[
t = t_0 +n\Delta T~~~,~~~x = x_0+i\Delta~~~,~~~y = y_0+j\Delta~~~,~~~z = z_0+k\Delta,
\]
where $n = 0,1,2, \dots N$, $i = 0,1,2, \dots I$, $j = 0,1,2, \dots J$, 
and $k = 0,1,2, \dots K$.  Next we employ the Alternating Direction Implicit (ADI) Method using the
Crank-Nicholson finite differencing scheme \cite{STRIKWERDA}. The ADI method is
especially useful in solving parabolic multidimensional equations on rectangular
grids. For problems with three spatial dimensions, such as Eq. (\ref{DIF_ONE}), 
the ADI method is implemented by splitting each time step of size $\Delta T$ 
into three steps of size $\Delta T/3$. At the current fractional time step,
$n+\athird$ for example,
only one of the spatial derivatives are evaluated and the others are
evaluated at the previous time step, $n$. If we choose not to split the time step
in such a way, the solution to Eq. (\ref{DIF_ONE}), after discretizing, would require us to invert a 
large matrix of the same size as our grid. The purpose of splitting the time step
in such a way is to reduce the problem in question into the solution of
a tridiagonal matrix (as shown below). To illustrate this, Eq. (\ref{DIF_ONE}) 
is discretized using the ADI method with the Crank-Nicholson finite differencing scheme. 
First discretize and evaluate the $x$ derivatives at the current time step $n+\athird$:

\begin{eqnarray}
\frac{u^{n+\athird}_{ijk} - u^n_{ijk}}{\Delta T/3} = 
\frac{A_n}{\Delta^2}\left[u^{n+\athird}_{i+1jk}+u^{n+\athird}_{i-1jk}
-2u^{n+\athird}_{ijk}
 +u^n_{ij+1k}+u^n_{ij-1k}\right.\nonumber\\
\left. -2u^{n}_{ijk}+
 u^n_{ijk+1}+u^n_{ijk-1}
-2u^{n}_{ijk}\right]\nonumber\\
+{\rm ~linear~terms}.
\label{STEP1}
\end{eqnarray}

Increment the time step by $1/3$ and evaluate the $y$ derivatives at current time step $n+\twothird$:

\begin{eqnarray}
\frac{u^{n+\twothird}_{ijk} - u^{n+\athird}_{ijk}}{\Delta T/3} = 
\frac{A_n}{\Delta^2}\left[u^{n+\athird}_{i+1jk}+u^{n+\athird}_{i-1jk}
-2u^{n+\athird}_{ijk}
 +u^{n+\twothird}_{ij+1k}+u^{n+\twothird}_{ij-1k}\right.\nonumber\\
\left. -2u^{n+\twothird}_{ijk}+
 u^{n+\athird}_{ijk+1}+u^{n+\athird}_{ijk-1}
-2u^{n+\athird}_{ijk}\right]\nonumber\\
+{\rm ~linear~terms}.
\label{STEP2}
\end{eqnarray}

Increment the time step again by $1/3$ and evaluate the $z$ derivatives at the current time step $n+1$:

\begin{eqnarray}
\frac{u^{n+1}_{ijk} - u^{n+\twothird}_{ijk}}{\Delta T/3} = 
\frac{A_n}{\Delta^2}\left[u^{n+\twothird}_{i+1jk}+u^{n+\twothird}_{i-1jk}
-2u^{n+\twothird}_{ijk}
 +u^{n+\twothird}_{ij+1k}+u^{n+\twothird}_{ij-1k}\right.\nonumber\\
\left. -2u^{n+\twothird}_{ijk}+
 u^{n+1}_{ijk+1}+u^{n+1}_{ijk-1}
-2u^{n+1}_{ijk}\right]\nonumber\\
+{\rm ~linear~terms}.
\label{STEP3}
\end{eqnarray}

In Eq.'s (\ref{STEP1}) to (\ref{STEP3}),  ``linear terms'' refers to the terms in (\ref{DIF_ONE}) which contain first
order derivatives. Since the Crank-Nicholson finite differencing scheme is stable only if the terms
with the highest order of derivatives are implicit, the lower order terms 
are allowed to be either explicit or implicit.  We assume  here for brevity in notation that 
the linear terms are explicit and evaluated at the previous time step. Therefore, we do not 
write the linear terms out.

The solution to Eq.'s (\ref{STEP1}) to (\ref{STEP3}) at every time step, is simply the solution of
a tridiagonal matrix at every $n$. For example, 
Eq. (\ref{STEP1}) can be arranged as follows:

\begin{equation}
\alpha^n u^{n+\athird}_{i-1jk}+\beta^n u^{n+\athird}_{ijk}+
\gamma^n u^{n+\athird}_{i+1jk} = \delta^n_{ijk},
\label{ABC}
\end{equation}

where
\[
\alpha^n = \gamma^n = -\frac{A_n\Delta T}{3\Delta^2}~~~,~~~
\beta^n = 2\frac{A_n\Delta T}{3\Delta^2}, 
\]
and

\[
\delta^n_{ijk} = \frac{A_n\Delta T}{3\Delta^2}\left[
 u^n_{ij+1k}+u^n_{ij-1k}
-2u^{n}_{ijk}+
 u^n_{ijk+1}+u^n_{ijk-1}
-2u^{n}_{ijk}\right]
+{\rm linear~terms}.
\]

Eq. (\ref{ABC}) is simply an equation of the form:
\begin{equation}
{\bf M}\cdot{\bf u} = {\bf d},
\label{MUD}
\end{equation}
where ${\bf M}$ is a tridiagonal matrix.  Since we know ${\bf M}$ and
${\bf d}$ at every previous time step $n$, we can in principal solve
for ${\bf u}$ at every current time step $n+\athird$, given the
boundary conditions $u^0_{ijk}$, $u^n_{0jk}$, $u^n_{i0k}$ and
$u^n_{ij0}$.  The inversion of sparse matrices (such as tridiagonal
matrices) is usually numerically trivial.

In solving Eq. (\ref{DIF_ONE}) one needs to specify the boundary
conditions suitable to solving a second order differential equation.
In this work we have specified $u^0_{ijk}$ by Eq. (\ref{idis}) and
required that $u$ vanish at the boundary of space: $u^n_{0jk}$,
$u^n_{i0k}$, $u^n_{ij0}$, $u^n_{Ijk}$, $u^n_{iJk}$ and $u^n_{ijK} =
0$. Furthermore we require that the first derivatives vanish at
$x,y,z\rightarrow \pm\infty$.  Therefore, $u^n_{Ijk}-u^n_{I-1jk}\sim
0$, $u^n_{1jk}-u^n_{0jk}\sim 0$, and so on for $j$ and $k$.


\begin{thebibliography}{99}

\bibitem{Mueller1}A. H. Mueller, {\em Nucl. Phys.} {\bf B572} 227 (2000).

\bibitem{Mueller2}A. H. Mueller, {\em Phys. Lett.} {\bf B475}, 220, (2000).

\bibitem{Geiger}K. Geiger, {\em Phys. Rep.} {\bf 258}, 237 (1995).

\bibitem{Karietal}
K. J. Eskola, K. Kajantie, P. V. Ruuskanen, and K. Tuominen, 
{\em Nucl. Phys.}  {\bf B570}, 379 (2000).

\bibitem{Wang}X.~Wang, {\em Phys. Rept.}  {\bf 280}, 287 (1997).

\bibitem{MV}L. McLerran and R. Venugopalan, {\em Phys. Rev.} {\bf D} 49, 
2233 (1994); {\em Phys. Rev.} {\bf D} 49, 3352 (1994); {\em Phys. Rev.}
 {\bf D} 50, 2225 (1994).

\bibitem{Larry}L. McLerran, private communication.

\bibitem{RajGavai}R. V. Gavai and R. Venugopalan, {\em Phys. Rev.}
{\bf D54} 5795 (1996).

\bibitem{JKMW}J. Jalilian-Marian, A. Kovner, L. McLerran, and H. Weigert, 
{\em Phys. Rev.} {\bf D} 55, 5414 (1997). 

\bibitem{Kovchegov}Y.~V.~Kovchegov, {\em Phys. Rev.}  {\bf D54}, 5463 (1996).

\bibitem{AJMV}A. Ayala, J. Jalilian-Marian, L. McLerran, and
R. Venugopalan, {\em Phys. Rev.}  {\bf D52}, 2935 (1995); {\it ibid.}
{\bf D53} 458 (1996).

\bibitem{JKLW}J. Jalilian--Marian, A. Kovner, L. McLerran, and
H. Weigert, {\em Phys. Rev.}  {\bf D55} (1997) 5414;
J. Jalilian-Marian, A. Kovner, A. Leonidov, and H. Weigert, {\em
Nucl. Phys.} {\bf B504} 415 (1997); J. Jalilian-Marian, A. Kovner, and
H. Weigert, {\em Phys. Rev.} {\bf D59} 014015 (1999); L. McLerran and
R. Venugopalan, {\em Phys. Rev.} {\bf D59} 094002 (1999).

\bibitem{KMW}A. Kovner, L. McLerran, and H. Weigert, {\em Phys. Rev.}
 {\bf D} 52, 6231 (1995); {\em Phys. Rev.} {\bf D} 52, 3809 (1995).  

\bibitem{PYM} Y. V. Kovchegov and D. H. Rischke, {\em Phys. Rev.}
{\bf C56} (1997) 1084; M. Gyulassy and L. McLerran, {\em Phys. Rev.}
{\bf C56} (1997) 2219; S. G. Matinyan, B. M\"uller and D. H. Rischke,
{\em Phys. Rev.} {\bf C56} (1997) 2191; {\em Phys. Rev.} {\bf C57}
(1998) 1927; Xiao-feng Guo, {\em Phys. Rev.} {\bf D59} 094017 (1999).


\bibitem{AlexRaj1}A. Krasnitz and R. Venugopalan, 
hep-ph/9706329, hep-ph/9808332, {\em Nucl. Phys.} {\bf B557} 237 (1999).

\bibitem{AlexRaj2}A. Krasnitz and R. Venugopalan,  {\em Phys. Rev. Lett.}  
{\bf 84}, 4309 (2000).

\bibitem{AlexRaj3}A. Krasnitz and R. Venugopalan, hep-ph/0007108, submitted 
to {\em Phys. Rev. Lett.}.

\bibitem{LifPit}E.M Lifshitz and L.P Pitaevskii, 
``Physical Kinetics'', Pergamon Press, (1981).

\bibitem{Raj99}R.~Venugopalan, {\em Acta Phys. Polon.} 
{\bf B30}, 3731 (1999).

\bibitem{BiroMuller}T. S. Biro, B. M\"uller and X-N. Wang, {\em Phys. Lett.}
 {283 \bf B}, 171 (1992); K. J. Eskola, B. M\"uller and X-N. Wang, {\em 
Phys. Lett.} {\bf 374B}, 21 (1996); L. Kadanoff and G. Baym, 
``Quantum Statistical Mechanics: Green's function methods in 
equilibrium and nonequilibrium problems'', 
Addison-Wesley Pub. Co., Advanced Book Program, (1989).

\bibitem{AMDScomm}R. Baier, A. H. Mueller, D. Schiff, D. T. Son, in progress; 
private communication by A. H. Mueller and D. T. Son.

\bibitem{BjKoSo}J.~D.~Bjorken, J.~B.~Kogut and D.~E.~Soper, {\em Phys. 
Rev.}  {\bf D3}, 1382 (1971).

\bibitem{Bjorken}
J.~D.~Bjorken, {\em Phys. Rev.}  {\bf D27}, 140 (1983); K. Kajantie and 
L. D. McLerran, {\em Nucl. Phys.} {\bf B214} (1983) 261; G. Baym et al., 
{\em Nucl. Phys.} {\bf A407} (1983) 541; M. Gyulassy and T. Matsui, {\em 
Phys. Rev.} {\bf D29} (1984) 419.

\bibitem{Baym}G. Baym, {\em Phys. Lett.} {\bf B138} 19 (1984).

\bibitem{ChapCowl}S. Chapman and T. G. Cowling, ``The mathematical theory 
of non--uniform gases'', Cambridge University Press, (1970).

\bibitem{PPVW}M. Prakash, M. Prakash, R. Venugopalan, and G. Welke, 
{\em Phys. Repts.} {\bf 227} (1993) 321.

\bibitem{DumitruNayak}G. C. Nayak, A. Dumitru, L. McLerran, and W. Greiner, 
hep-ph/0001202; S.A. Bass, A. Dumitru, nucl-th/0001033.

\bibitem{Baymetal}G.~Baym, H.~Monien, C.~J.~Pethick and 
D.~G.~Ravenhall, {\em Phys. Rev. Lett.}  {\bf 64}, 1867 (1990).

\bibitem{LeBellac}M. Le Bellac, ``Thermal Field Theory'', Cambridge University 
Press, (1996).

\bibitem{SGHASYB}A.~Selikhov and M.~Gyulassy,
{\em Phys. Lett.}  {\bf B316}, 373 (1993); H.~Heiselberg,
{\em Phys. Rev. Lett.}  {\bf 72}, 3013 (1994); 
P.~Arnold, D.~T.~Son and L.~G.~Yaffe,
{\em Phys. Rev.}  {\bf D59}, 105020 (1999); D.~Bodeker, 
{\em Phys. Lett.}  {\bf B426}, 351 (1998).

\bibitem{EskKajRuu}K. J. Eskola, K. Kajantie, and P. V. Ruuskanen, 
{\em Nucl. Phys.} {\bf B323} (1989) 37.

\bibitem{Gavin}S. Gavin, {\em Nucl. Phys.} {\bf B351} (1991) 561.

\bibitem{HX1}H. Heiselberg and X. N. Wang, {\em Phys. Rev.} {\bf C53} 
(1996) 1892.

\bibitem{HX2}H. Heiselberg and X. N. Wang, {\em Nucl. Phys.} {\bf B462} 
(1996) 389.

\bibitem{AdrianMiklos}A. Dumitru and M. Gyulassy, hep-ph/0006257. 

\bibitem{BiroDoorn}
T.~S.~Biro, E.~van Doorn, B.~Muller, M.~H.~Thoma and X.~N.~Wang,
{\em Phys. Rev.}  {\bf C48}, 1275 (1993); L.~Xiong and E.~Shuryak,
{\em Phys. Rev.}  {\bf C49}, 2203 (1994); D.~M.~Elliott and D.~H.~Rischke,
{\em Nucl. Phys.}  {\bf A671}, 583 (2000).

\bibitem{JeffRaj}J. Bjoraker and R. Venugopalan, in preparation.

\bibitem{Recipes}W. H. Press, B. P. Flannery, S. A. Teukolsky, and 
W. T. Vetterling, ``Numerical Recipes'', Cambridge University Press, (1989).

\bibitem{STRIKWERDA}J. C. Strikwerda,
``Finite Difference Schemes and Partial Differential Equations'',
Wadsworth \& Brooks, (1989).



\end{thebibliography}
\end{document}